\documentclass[aps,reprint,superscriptaddress,showpacs,nofootinbib]{revtex4-2}

\usepackage{amsmath,amsthm,amssymb}
\usepackage{dcolumn,bm,hyperref}
\usepackage{graphicx,subfigure,verbatim}
\usepackage{txfonts}    
\usepackage{xcolor}
\usepackage[export]{adjustbox}

\renewcommand{\vec}[1]{{\bm{\mathrm{#1}}}}
\newcommand{\vhat}[1]{\hat{\bm{\mathrm{#1}}}}

\let\Im\undefined
\DeclareMathOperator{\Im}{Im}

\newcommand{\bra}[1]{\langle{#1}|}
\newcommand{\ket}[1]{|{#1}\rangle}

\usepackage{xcolor}

\begin{document}
\title{Microscopic study of orbital textures}

\author{Seungyun Han}
\affiliation{Department of Physics, Pohang University of Science and Technology, Pohang 37673,Korea}

\author{Hyun-Woo Lee}
\email{hwl@postech.ac.kr}
\affiliation{Department of Physics, Pohang University of Science and Technology, Pohang 37673,Korea}

\author{Kyoung-Whan Kim}
\email{kwk@kist.re.kr} 
\affiliation{Center for Spintronics, Korea Institute of Science and Technology, Seoul 02792, Korea}

\pacs{\quad}

\date{\today}

\begin{abstract}
Many interesting spin and orbital transport phenomena originate from orbital textures, referring to $\vec{k}$-dependent orbital states. Most of previous works are based on symmetry analysis to model the orbital texture and analyze its consequences. However the microscopic origins of orbital texture and its strength are largely unexplored. In this work, we derive the orbital texture Hamiltonians from microscopic tight-binding models for various situations. To form an orbital texture, $\vec{k}$-dependent hybridization of orbital states are necessary. We reveal two microscopic mechanisms for the hybridization: (i) lattice structure effect and (ii) mediation by other orbital states. By considering the orbital hybridization, we not only reproduce the orbital Hamiltonian obtained by the symmetry analysis but also reveal previously unreported orbital textures like orbital Dresselhaus texture and anisotropic orbital texture. The orbital Hamiltonians obtained here would be useful for analyzing the orbital physics and designing the materials suitable for spin-orbitronic applications. We show that our theory also provides useful microscopic insight into physical phenomena such as the orbital Rashba effect and the orbital Hall effect. Our formalism is so generalizable that one can apply it to obtain effective orbital Hamiltonians for arbitrary orbitals in the presence of periodic lattice structures. 
\end{abstract}

\maketitle

\section{Introduction}
Spin-momentum couplings induce spin eigendirections to vary with $\vec{k}$. Such $\vec{k}$-space spin textures generate many interesting spin phenomena, such as the spin Hall effect~\cite{sinova2004,sinova2015} and spin-orbit torque~\cite{miron2011,liu2012,kurebayashi2014}, and thus provide a useful starting point to analyze spin phenomena. Considering that spin-momentum couplings are possible only when either time-reversal symmetry (TRS) or inversion symmetry (IS) is broken~\cite{sinova2004}, a nontrivial $\vec{k}$-space spin texture is possible only when at least one of the two symmetries is broken. 

Recently, many studies in the field of spintronics have been expanded to utilize the orbital degree of freedom of electrons~\cite{bernevig2005,go2018,jo2018,choi2021,bhowal2021,cysne2021}. The orbital degree of freedom usually generates larger responses to external perturbations than the spin degree of freedom does, since the orbital energy scale is determined by the crystal field and larger than the energy sace of the spin-orbit coupling that governs the spin dynamics. In addition, since electron orbital carries angular momentum larger than that of spin ($=\hbar/2$), it is expected to transfer angular momentum more efficiently.

\begin{figure}[]
\centering
\includegraphics[width=0.3\textwidth]{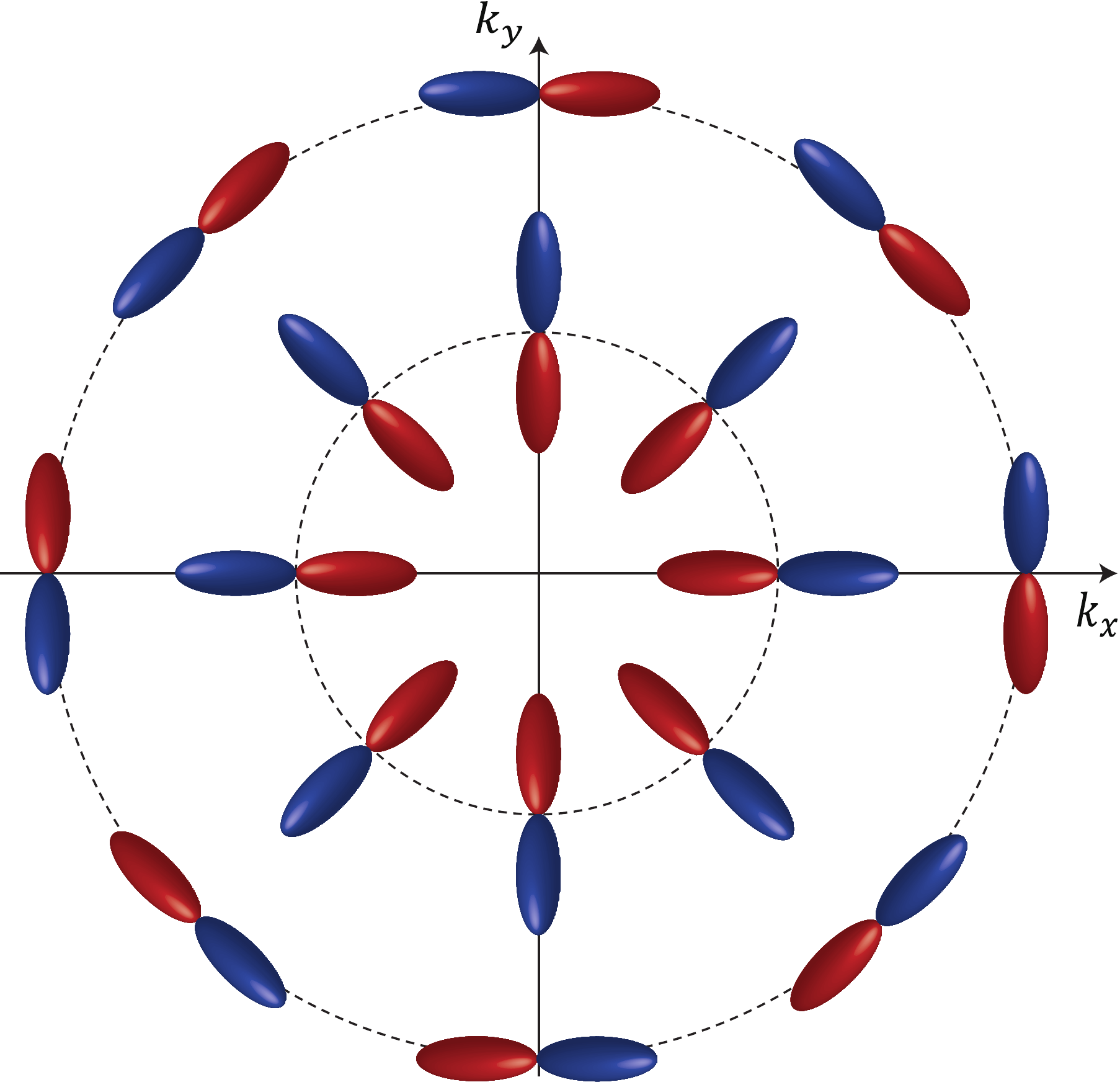}
\caption{\label{Fig:2drt} The radial and tangential orbital texture in 2D}
\end{figure}

Many orbital physics start with the generation of an orbital current from the so-called orbital texture. An orbital texture system is a system whose orbital eigenstates vary with $\vec{k}$. Unlike spin, non-trivial orbital texture is formed even with both TRS and IS~\cite{tokatly2010,go2018,han2022}. A widely used model is so-called the radial-tangential ($r$-$t$) $p$-orbital model. A two-dimensional (2D) $r$-$t$ Hamiltonian is presented in Eq.~(\ref{otHamiltonian}) as the simplest $r$-$t$ model~\cite{tokatly2010,go2018,han2022}. In the ($p_x$, $p_y$) basis, the Hamiltonian of the model reads,
\begin{eqnarray} \label{otHamiltonian}
\mathcal{H}=\frac{\hbar^2k^2}{2m}+ \frac{\hbar^2\eta}{2m}\begin{pmatrix}
k_x^2 & k_xk_y\\
k_xk_y & k_y^2
\end{pmatrix},
\end{eqnarray}
where $k=|\vec{k}|$. The eigenstates of $\mathcal{H}$ in Eq.~(\ref{otHamiltonian}) are $p_r=(\cos\theta_\vec{k},\sin\theta_\vec{k})$ and $p_t=(-\sin\theta_\vec{k},\cos\theta_\vec{k})$, where $\theta_\vec{k}=\arg(k_x+ik_y)$. $p_r$ and $p_t$ are called radial and tangential $p$ states, respectively. Equation~(\ref{otHamiltonian}) results in the orbital texture in Fig.~\ref{Fig:2drt}.

The inner band in Fig.~\ref{Fig:2drt} is called the radial $p$-orbital band since the real orbital of its eigenstate is aligned along the momentum direction and the outer band in Fig.~\ref{Fig:2drt} is called the tangential $p$-orbital band. The orbital texture in Fig.~\ref{Fig:2drt} generally exists in many $p$-orbital systems~\cite{go2018,jo2018,tokatly2010,baek2021}, and it generates many orbital physics phenomena like the orbital Hall effect (OHE)~\cite{go2018,tokatly2010,jo2018,choi2021}. The energy difference between the radial and tangential orbitals characterizes the strength of the orbital texture and is parameterized by $\eta$ in Eq.~(\ref{otHamiltonian}). However, there is no comprehensive understanding of the microscopic origin of the orbital texture because the microscopic origin of $\eta$ is studied in only for limited cases~\cite{wu2008,kim2019,go2018,ko2020} and most studies rely on the symmetry argument on the existence of the $r$-$t$ orbital texture~\cite{tokatly2010,han2022}. Therefore, a comprehensive microscopic study of orbital textures is desired.

The aim of this article is to systematically study the microscopic origins of orbital textures starting from tight-binding models for various physical situations. Our approach for deriving orbital textures goes beyond the symmetry analysis and enables investigating orbital textures in various lattice systems with various orbitals. In this paper, we focus on orbital textures driven by hybridization of orbitals with the same orbital quantum number ($l$), since these are the illustrative systems with interesting orbital physics like the OHE. For example, for the orbital texture in Fig.~\ref{Fig:2drt}, $p_x$ and $p_y$ (for $l=1$) are hybridized. Accordingly, we reveal two mechanisms to generate orbital textures: (i) hybridization through the lattice structure and (ii) hybridization mediated by another orbitals with different $l$. For the mechanism (i), the lattice structure mixes the orbitals in same $l$ (e.g., $p_x$ and $p_y$) while the mechanism (ii) corresponds to orbitals in the same $l$ effectively mixed by another orbital in different $l$ (e.g., $sp$ hybridization~\cite{go2018}).
 
\begin{figure}
	\centering
	\includegraphics[width=0.4\textwidth]{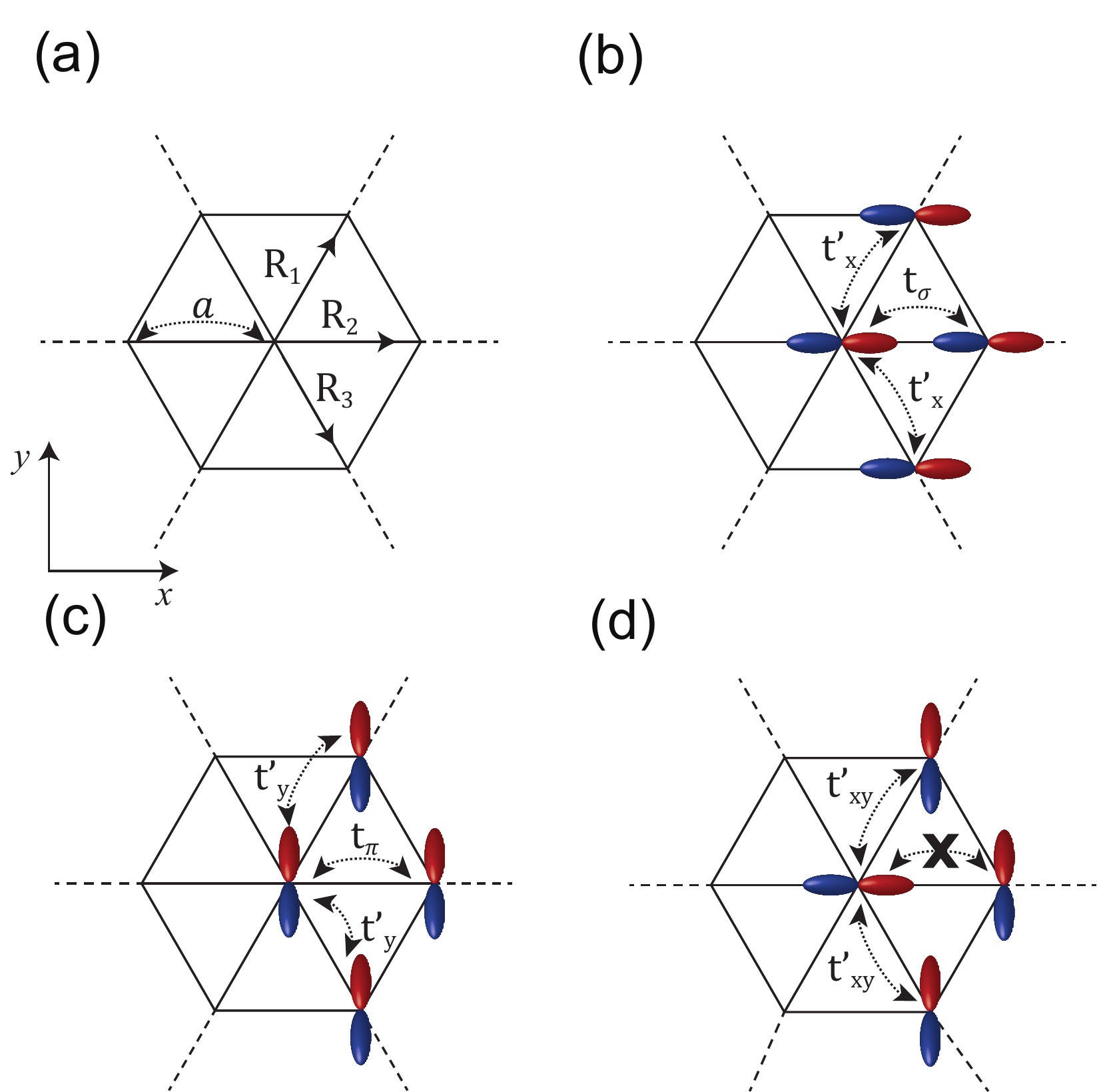}
 	\caption{\label{Fig:ptriagnular} (a) Vectors of NN. $a$ denotes the lattice constant. (b) Hopping rules for $p_x$ orbital. Along the $R_2$ direction, the hopping amplitude is $t_\sigma$, and along the $R_1$ and $R_3$ directions, the hopping amplitude is $t_x'=(t_\sigma+3t_\pi)/4$. (c) Hopping rules for $p_y$ orbital. Along the $R_2$ direction, hopping amplitude is $t_\pi$ and along the $R_1$ and $R_3$ directions, the hopping amplitude is $t_y'=(t_\pi+3t_\sigma)/4$.	(d) Hopping rules for hopping between $p_x$ and $p_y$. Along the $R_2$ direction, hopping amplitude is zero and, along the $R_1$ and $R_3$ directions, the hopping amplitude is $t_{xy}'=\sqrt{3}(t_\sigma-t_\pi)/4$.}
\end{figure}
 
 
The rest of the paper is organized as follows. In Sec.~\ref{Sec:basic models}, we examine simple models for the two mechanisms. We derive the orbital texture of a triangular lattice with the nearest-neighbor (NN) hopping and a square lattice with next-nearest-neighbor (NNN) hopping as example systems of the mechanism (i), and square and cubic lattices with $sp$ hybridization as example systems of the mechanism (ii). In Sec.~\ref{Sec:advanced models}, more generalized models are investigated. We see how the orbital textures are modified when one considers other orbitals like $d$ orbitals. Also, we consider various lattice structures, for more realistic situations,
such as a hexagonal lattice, a bilayer square lattice, a multilayer thin film, face-centered cubic (FCC) structure, and body-centered cubic (BCC) structure. 
In Sec.~\ref{Sec:discussion}, we discuss how the orbital Hall conductivity depends on the types of the microscopic features of the orbital texture. 
In Sec.~\ref{Sec:summary}, we summarize the paper.


\section{Basic models\label{Sec:basic models}}

In this section, for simplicity, we derive $p$-orbital textures like Fig.~\ref{Fig:2drt}. We investigate the simplest model for each of the mechanisms (i) and (ii). For the mechanism (i), we do not consider $s$ and $d$ orbitals since the lattice structure drives system to have an orbital texture even without hybridization with an orbital with another $l$. 
For the simplest model, we adopt a triangular lattice with the NN hopping and a square lattice with the NNN hopping which mixes $p_x$ and $p_y$ orbitals. For the mechanism (ii), we derive orbital-texture Hamiltonians on square and cubic lattices with $s$ and $p$ orbitals. Up to the NN hopping, there is no direct hybridization between $p_x$ and $p_y$ orbitals but $p_x$ and $p_y$ orbitals are effectively mixed through the $sp$ hybridization. Lastly, we deal with a triangular lattice 
in which two mechanisms work together and show that the two mechanisms give linearly addable contributions to $\eta$ up to the lowest order.

\subsection{Mechanism (i): Orbital textures driven by the lattice structure}

\subsubsection{2D Triangular lattice with $p$-orbitals}

In a 2D lattice in the $xy$ plane with the mirror symmetry, the $p_z$ orbital is decoupled from the $p_x$ and $p_y$ orbitals and thus does not contribute to the orbital texture formation. Therefore, we ignore the $p_z$ orbital for simplicity throughout this paper unless specified. To write down the tight-binding model for the 2D triangular lattice with NN hopping, we adopt the NN hopping rules in Fig.~\ref{Fig:ptriagnular}. 
The anisotropic orbital hopping in Fig.~\ref{Fig:ptriagnular} generates $\vec{k}$-dependent orbital eigenstates (i.e., the orbital texture). The tight-binding Hamiltonian in the ($p_x$, $p_y$) basis is given by
\begin{eqnarray} \label{triangularporbital}
\mathcal{H} = \begin{pmatrix}
H_{p_xp_x} & H_{p_xp_y}\\
H_{p_xp_y} & H_{p_yp_y}
\end{pmatrix}, 
\end{eqnarray}
where $H_{p_xp_x/p_yp_y}=3t_{\pi/\sigma}\cos(ak_x/2)\cos(\sqrt{3}ak_y/2)
+t_{\sigma/\pi}[\cos(ak_x/2)^2+\cos(ak_x/2)\cos(\sqrt{3}ak_y/2)-2\sin(ak_x/2)^2]$, $H_{p_xp_y}=H_{p_yp_x} = \sqrt{3}(t_\pi-t_\sigma)\sin(ak_x/2)\sin(\sqrt{3}ak_y/2)$, and $a$ is the lattice constant. To derive the effective Hamiltonian near the $\Gamma$ point, we expand the Hamiltonian up to $k^2$.
\begin{equation} \label{triangularHamiltonian}
\mathcal{H} = 3(t_\sigma+t_\pi)\left(1-\frac{a^2k^2}{4}\right) - \frac{3a^2}{4}(t_\sigma-t_\pi)\begin{pmatrix}
k_x^2 & k_xk_y \\
k_xk_y & k_y^2
\end{pmatrix}.
\end{equation}
Note that Eq.~(\ref{triangularHamiltonian}) is equivalent to the 2D $r$-$t$ Hamiltonian [Eq.~(\ref{otHamiltonian})] for $\eta = (t_\sigma-t_\pi)/(t_\sigma+t_\pi)$. The strength of the orbital texture is determined by the difference between $\sigma$ and $\pi$ hopping amplitudes which measures the orbital hopping anisotropy. Further, the dispersion is isotropic in $\vhat{k}$, i.e., there is no Fermi surface warping in this model. This is because the triangular lattice has $\pi/3$-rotation symmetry, implying the warping term $\propto k^3$ which is neglected in our calculation. Therefore, a triangular lattice is an ideal system which has Eq.~(\ref{otHamiltonian}) as an effective Hamiltonian near the $\Gamma$ point.

\begin{figure}[]
	\centering
	\includegraphics[width=0.45\textwidth]{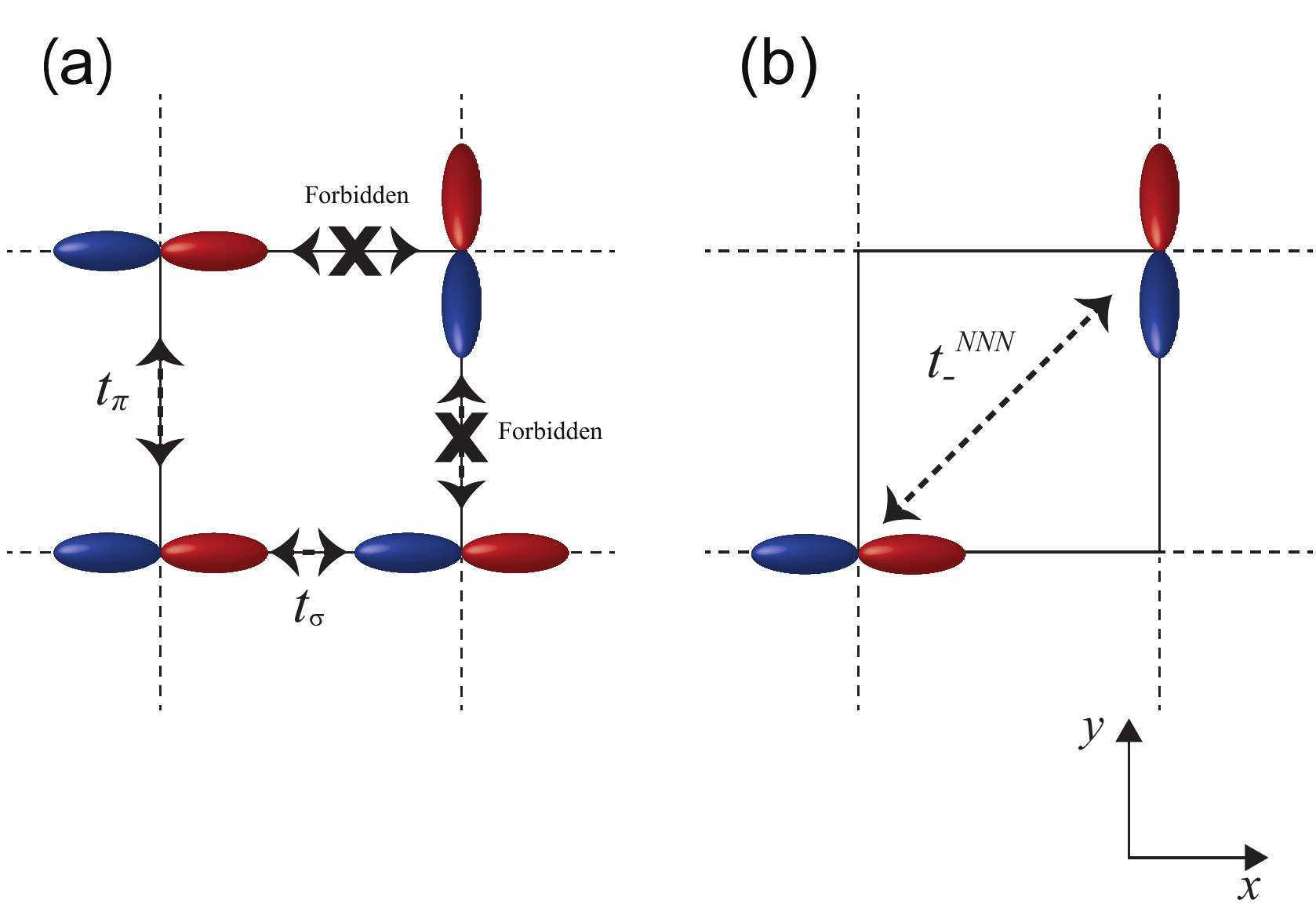}
	\caption{\label{Fig:squarephopping} $p$-orbital hopping rules in a square lattice. (a) Hopping rules for NN hoppings. $p_x$ and $p_y$ orbitals are not hybridized. (b) Hopping rules for NNN hopping. $t_-^{\rm NNN}$ is defined in the main text. The same coordinate system is used for square lattices below. }
\end{figure}


\subsubsection{Square lattice with next nearest neighbor hopping}

\begin{figure}
\centering
\includegraphics[width=0.53\textwidth]{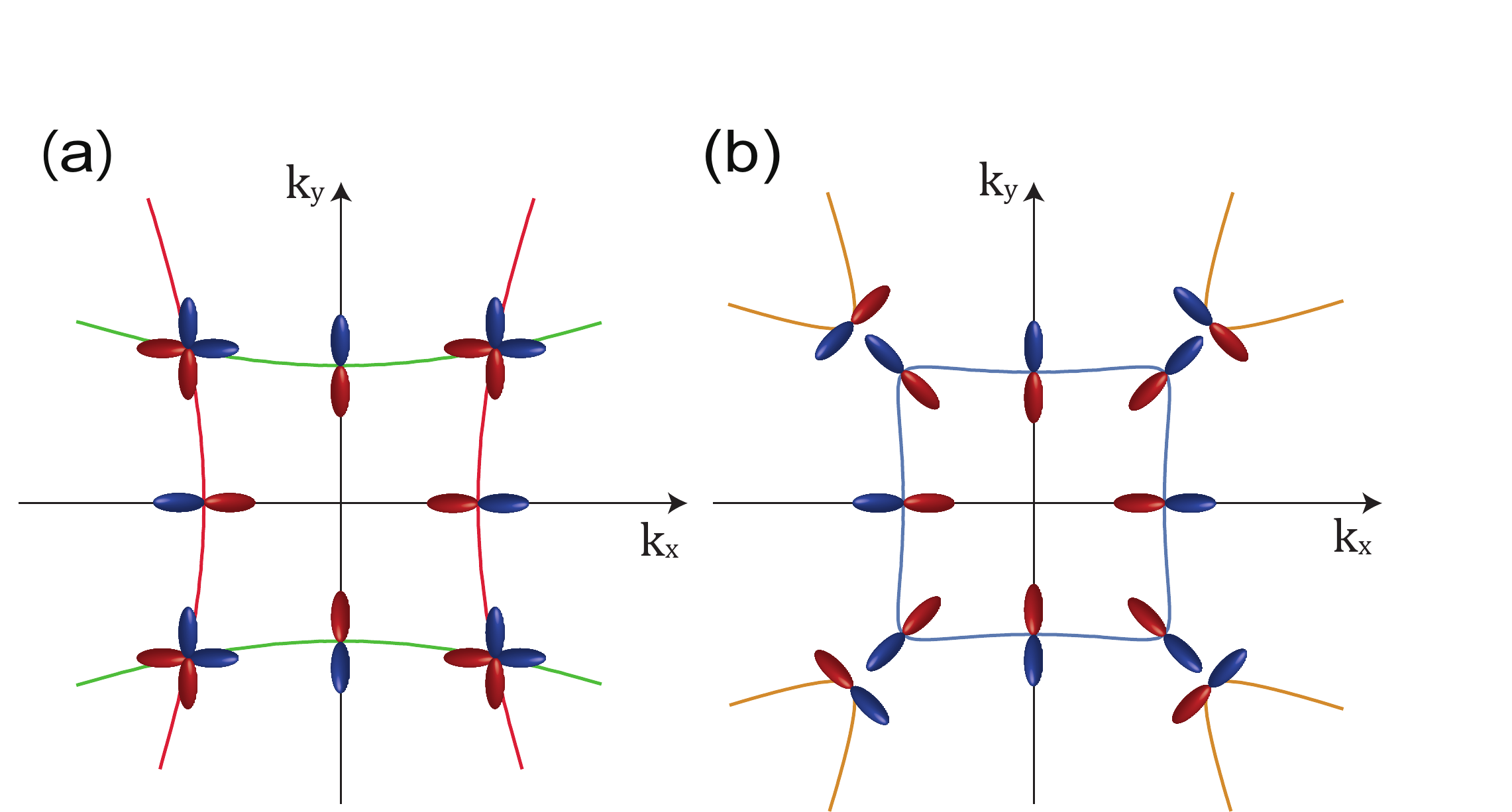}
\caption{\label{Fig:otsquare} Fermi surface and $p$-orbital character for a square lattice. (a) Up to the NN hopping, there is no $p$-orbital texture. (b) $p$-orbital texture is formed when the NNN hopping are taken into account.}
\end{figure}

Next, we consider a square lattice with $p$ orbitals up to the NNN hopping. In a square lattice, $p_x$ and $p_y$ orbitals are not directly mixed up to the NN hopping [Fig.~\ref{Fig:squarephopping}(a)], resulting in no orbital texture. However, $p_x$ and $p_y$ orbitals can be hybridized via NNN hopping channels [Fig.~\ref{Fig:squarephopping}](b)]. This may generates an orbital texture. We now analytically write the Hamiltonian. First, the Hamiltonian up to the NN hopping is given by
\begin{align}\label{HpNN}
\mathcal{H}^{\rm NN}_p&= \begin{pmatrix}
H_{p_xp_x} & 0 \\
0 & H_{p_yp_y} 
\end{pmatrix} \nonumber \\
&\approx E_p^0+2(t_\sigma+t_\pi) -a^2\begin{pmatrix}
t_\sigma k_x^2 +t_\pi  k_y^2 & 0 \\
0& t_\pi k_x^2 + t_\sigma  k_y^2
\end{pmatrix},
\end{align}
where, in the second line, we expanded $H_{p_xp_x/p_yp_y} = E_p^0 + 2t_{\sigma/\pi} \cos(ak_x)+ 2t_{\pi/\sigma} \cos(ak_y)$ up to $k^2$. Since there is no direct mixing between $p_x$ and $p_y$ orbitals, Eq.~(\ref{HpNN}) does not exhibit an orbital texture. However, with the NNN hopping, there arises a hopping channel between $p_x$ and $p_y$ orbitals [Fig.~\ref{Fig:squarephopping}(b)]. Specifically, the NNN hopping term ($\mathcal{H}_p^{\rm NNN}$) is given by, up to $k^2$ order,
\begin{equation}
\mathcal{H}^{\rm NNN}_p= (4-2a^2k^2)t_+^{\rm NNN}
-4t_-^{\rm NNN}a^2\begin{pmatrix}
0& k_xk_y\\
k_xk_y & 0
\end{pmatrix},\label{HpNNN}
\end{equation}
where $t_\pm^{\rm NNN}=(t_{\sigma}^{\rm NNN}\pm t_{\pi}^{\rm NNN})/2$, and $t^{\rm NNN}_\sigma$ and $t^{\rm NNN}_\pi$ are the NNN-$\sigma$ and NNN-$\pi$ hopping integrals, respectively. The total Hamiltonian is then $\mathcal{H} = \mathcal{H}_p^{\rm NN}+\mathcal{H}_p^{\rm NNN}$. Now we discuss the role of $\mathcal{H}_p^{\rm NNN}$ more specifically. We plot the Fermi surface of the Hamiltonian without the $\mathcal{H}_p^{\rm NNN}$ term in Fig.~\ref{Fig:otsquare}(a), where the red and green bands have $p_x$ and $p_y$ orbital characters, respectively. They cross at the $k_x=\pm k_y$ points and gap is not opened since there is no mixing term between the $p_x$ and $p_y$ bands. Therefore, the orbital characters of the two bands do not change and there is no orbital texture in this system. However, with $\mathcal{H}_p^{\rm NNN}$, $p_x$ and $p_y$ orbitals are mixed and thus degeneracies are lifted at $k_x=\pm k_y$ points [Fig.~\ref{Fig:otsquare}(b)]. Accordingly, inner and outer bands are well separated having orbital texture. Near the $\Gamma$ point, this orbital texture resembles the $r$-$t$ type with a warped Fermi surface.\footnote{Different from the triangular case, a square lattice has the $\pi/2$ rotation symmetry and thus the Fermi surface warping term can appear even in the $k^2$ order.}

\subsection{Mechanism (ii): Orbital texture driven by hybridization between orbitals with different $l$}

\subsubsection{Square lattice with $sp$ orbitals}

\begin{figure}[]
	\centering
	\includegraphics[width=0.45\textwidth]{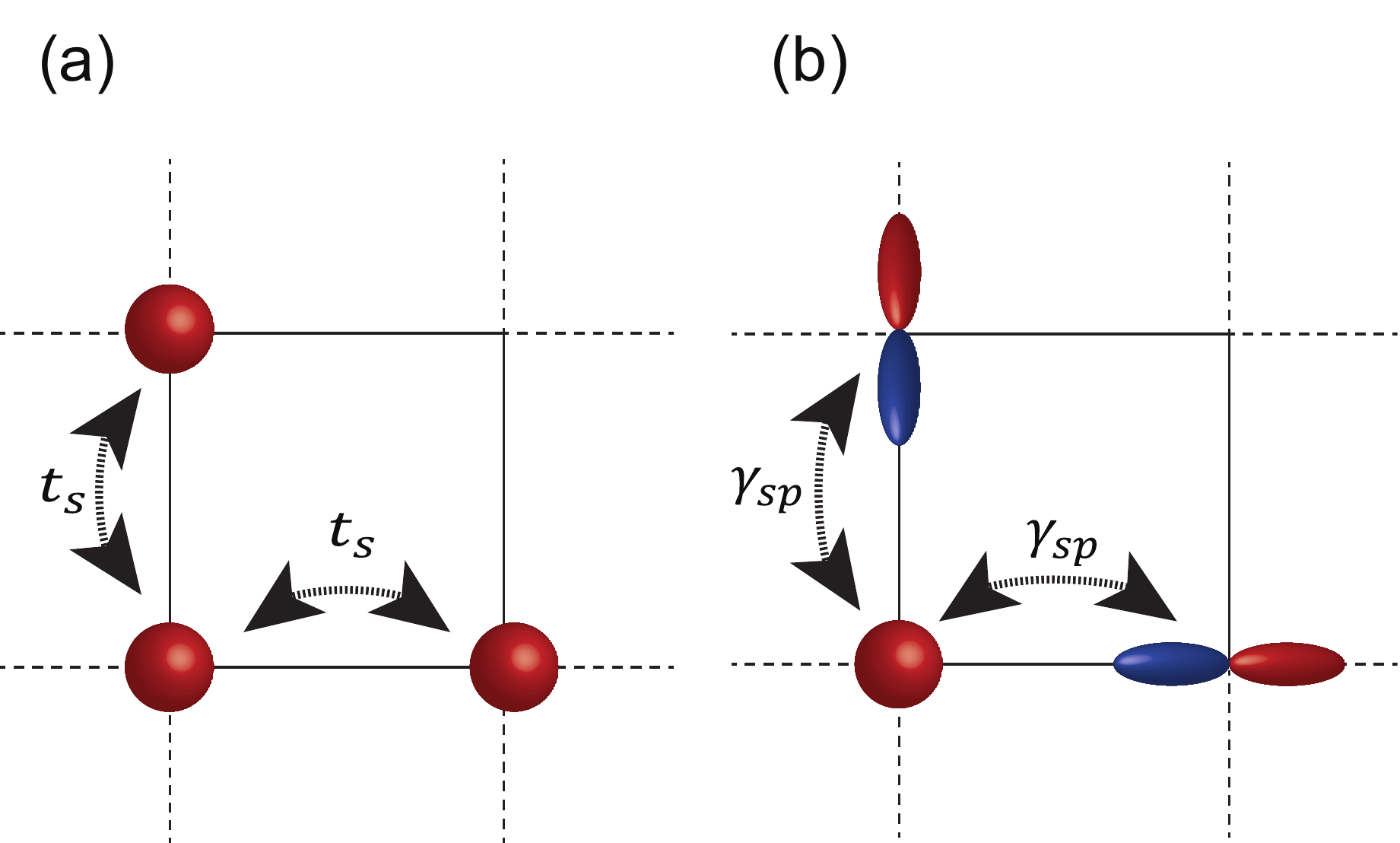}
	\caption{\label{Fig:sphopping} Hopping rules for $s$ orbital and $sp$ orbital. (a) $s$ orbital hopping. (b) Hopping rules for the hopping between $s$ and $p$ orbitals.}
\end{figure}

In this subsection, we derive the effective Hamiltonian of a square lattice with $sp$ orbitals up to the NN hopping including the $sp$ hybridization. 
As seen from the previous subsection, a square-lattice system does not have an orbital texture up to the NN hopping [Fig.~\ref{Fig:otsquare}(a)]. However, the $p_x$ and $p_y$ orbitals can be hybridized with $s$ orbital [Fig.~\ref{Fig:sphopping}(b)] which generates an effective hopping channel between the $p_x$ and $p_y$ orbitals (via the $sp$ hybridization) and allows a $p$-orbital texture to be formed. 
Using the hopping rules in Fig.~\ref{Fig:sphopping}, the Hamiltonian up to the NN hopping in the ($s$, $p_x$, $p_y$) basis is given by
\begin{equation} \label{sptightbinding}
\mathcal{H}=\begin{pmatrix}
H_{ss} & 2i\gamma_{sp}\sin(ak_x) & 2i\gamma_{sp}\sin(ak_y) \\
-2i\gamma_{sp}\sin(ak_x) &H_{p_xp_x} & 0 \\
-2i\gamma_{sp}\sin(ak_y) &0& H_{p_yp_y}
\end{pmatrix},
\end{equation}
where $H_{ss}= E_s^0 + 2t_s[\cos(ak_x)+\cos(ak_y)]$, $H_{p_xp_x/p_yp_y} = E_p^0 + 2t_{\sigma/\pi} \cos(ak_x)+ 2t_{\pi/\sigma} \cos(ak_y)$, $t_s$ is the hopping integral for the $s$ orbital, $\gamma_{sp}$ is strength of the $sp$ hybridization, and $E_s^0$ and $E_p^0$ are the on-site energies of $s$ and $p$ orbitals, respectively.

In Eq.~(\ref{sptightbinding}), we can see that the $sp$ hybridization mediates the mixing between the $p_x$ and $p_y$ orbitals. We show this point in two ways. First, without the $sp$ hybridization, $\gamma_{sp}=0$, the Fermi surface of $p$-orbital Hamiltonian is given Fig.~\ref{Fig:otsquare}(a); there is no orbital texture and two orbitals are degenerate at $k_x = \pm k_y$ points. However, with the nonzero $sp$ hybridization, this degeneracy is lifted. For example at $k_x=k_y=k/\sqrt{2}$, Hamiltonian [Eq.~(\ref{sptightbinding})] in the [$s$, $(p_x+p_y)/\sqrt{2}$, $(p_x-p_y)/\sqrt{2}$] basis is given by
\begin{equation} \label{squaresptransfomred}
\mathcal{H} =
\begin{pmatrix}
E_s-a^2t_sk^2& 2ia\gamma_{sp}k & 0\\
-2ia\gamma_{sp}k & E_p-\frac{a^2}{2}(t_\sigma +t_\pi)k^2 & 0\\
0 & 0&  E_p-\frac{a^2}{2}(t_\sigma +t_\pi)k^2
\end{pmatrix},
\end{equation}
up to $k^2$ order. As manifested in Eq.~(\ref{squaresptransfomred}), $(p_x+p_y)/\sqrt{2}$ interacts with $s$ orbital which lifts the degeneracy [Fig.~\ref{Fig:otsquare}(b)] while $(p_x-p_y)/\sqrt{2}$ does not. Therefore, the $sp$ hybridization plays a similar role as the $H_p^{\rm NNN}$ term in Eq.~(\ref{HpNNN}) in that it effectively mixes the $p_x$ and $p_y$ orbitals forming orbital texture.

\begin{figure}
\centering
\includegraphics[width=0.52\textwidth]{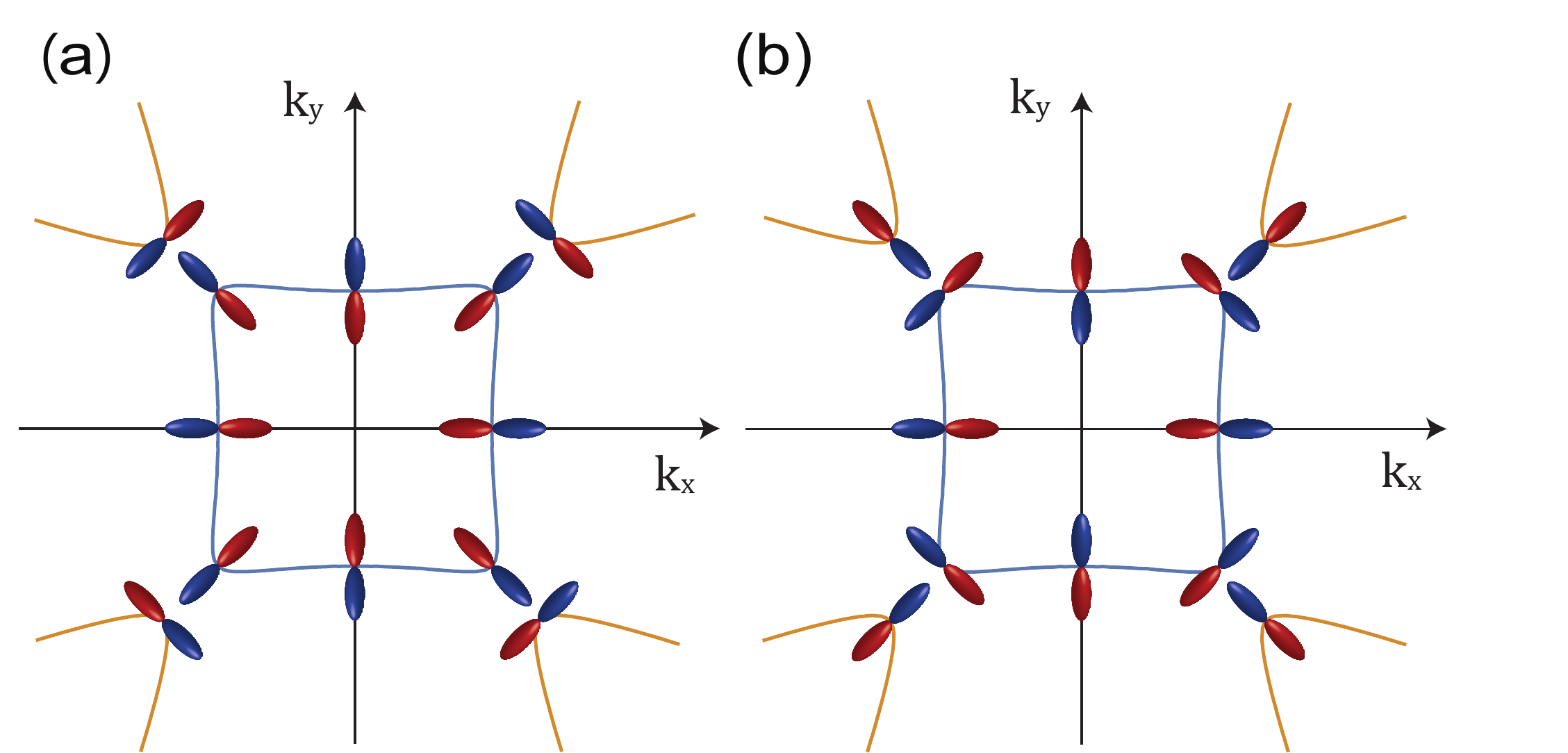}
\caption{\label{Fig:rashbadressel}{$r$-$t$ and Dresselhaus type orbital texture. (a) $r$-$t$ type texture of $p$ orbital. $p$ character rotates counterclockwise as $(k_x,k_y)$ is varied counterclockwise. (b) The Dresselhaus type of $p$ orbital. $p$ character rotates clockwise.} 
}
\end{figure}

Secondly, this can be seen more directly by using the L\"owdin downfolding technique \cite{lowdin1951}. The L\"owdin downfolding is a unitary transformation technique which divides the Hilbert space into two decoupled subspaces making the Hamiltonian block diagonal. Using this, we can get an effective projected Hamiltonian in a subspace of interest. The L\"owdin downfolding technique is briefly reviewed in the Appendix~\ref{Sec(A):Lowdin}. We derive the effective Hamiltonian for the $p$ bands by projecting the Hamiltonian [Eq.~(\ref{sptightbinding})] into the $p$-orbital subspace. According to Appendix~\ref{Sec(A):Lowdin}, we obtain up to $k^2$ order,
\begin{eqnarray} \label{squareporbital}
    \mathcal{H}_p=\mathcal{H}_p^{NN} + \frac{4a^2\gamma_{sp}^2}{E_p-E_s} \begin{pmatrix}
    k_x^2 & k_xk_y \\
    k_xk_y & k_y^2
    \end{pmatrix}.
\end{eqnarray}
The term proportional to $\propto \gamma_{sp}^2$ in Eq.~(\ref{squareporbital}) describes effective hopping between the $p_x$ and $p_y$ orbitals, mediated by the $s$ orbital. Note that the off-diagonal components lead to the orbital texture. One can deduce another implication from Eq.~(\ref{squareporbital}). Consider the orbital character of the inner band at $k_x = k_y >0$. If $t_\sigma + t_\pi > 0 $, the two bands have quadratic dispersions with negative effective masses. Then, due to the $sp$ hybridization, $p_x+p_y$ states have additional energy $\propto 4a^2\gamma_{sp}^2k^2/(E_p-E_s)$ while $p_x-p_y$ state does not. Then whether the additional energy is positive or negative determines the inner-band orbital character. Specifically, since $4a^2\gamma_{sp}^2k^2$ is always positive, the sign is solely determined by the difference of $E_p$ and $E_s$. As a result, the inner-band orbital character is $(p_x + p_y)/\sqrt{2}$ when $E_p-E_s < 0$ and $(p_x-p_y)/\sqrt{2}$ when $E_p-E_s >0$. Accordingly, the $r$-$t$ type orbital texture is formed when $E_p-E_s <0$ [Fig.~\ref{Fig:rashbadressel}(a)] while a \textit{Dresselhaus}-like orbital texture is formed when $E_p-E_s >0$ [Fig.~\ref{Fig:rashbadressel}(b)]\footnote{When $t_\sigma+t_\pi<0$, the $r$-$t$ type texture is formed for $E_p-E_s>0$ and the $Dresselhaus$ type texture is formed for $E_p-E_s <0$.}.

\subsubsection{Cubic lattice with $sp$}
As a direct generalization to three dimensions (3D), we can build an effective Hamiltonian of a cubic lattice with the $sp$ hybridization. For this case, we include the $p_z$ orbital with the same on-site energy as the $p_x$ and $p_y$ orbitals.\footnote{This is a good approximation for a bulk system. We relax this assumption in Sec.~\ref{Sec:advanced models} to consider thin films.} Similar to the 2D cases, we downfold the total Hamiltonian into the $p$-orbital subspace and obtain
\begin{align} \label{cubicsp}
\mathcal{H}_p &= E_p - a^2t_\pi k^2 -a^2(t_\sigma-t_\pi) \begin{pmatrix}
k_x^2  & 0 & 0 \\
0 &  k_y^2& 0 \\
0& 0 & k_z^2
\end{pmatrix} \nonumber \\
&\quad + \frac{4a^2\gamma_{sp}^2}{E_p-E_s}\begin{pmatrix}
k_x^2 &k_xk_y& k_xk_z\\
k_xk_y & k_y^2 & k_yk_z\\
k_xk_z & k_yk_z & k_z^2
\end{pmatrix},
\end{align}
which can be expressed in terms of the OAM operators $L_i$ 
\begin{align}\label{cubicsp2}
 \mathcal{H}_p &= E_p-\left(t_\sigma -\frac{4\gamma_{sp}^2}{E_p-E_s}\right)a^2k^2
-\frac{4a^2\gamma_{sp}^2}{E_p-E_s}(\vec{L}\cdot \vec{k})^2   \nonumber \\
&\quad +a^2(t_\sigma-t_\pi) \sum_ik_i^2L_i^2,
\end{align}
where the summation indices $i,j$ run over $i,j=x,y,z$ through out this paper. Also, through out this paper, we consider the dimensionless OAM operators (by dividing them by $\hbar$). When $t_\sigma = t_\pi$, the Hamiltonian has the radial and tangential orbitals as its eigenstates, similar to the triangular lattice model while the radial orbital now becomes 3D, $p_r = \hat{k_x}p_x + \hat{k_y}p_y + \hat{k_z}p_z$, and two tangential bands are degenerate as previously studied in \cite{go2018,park2022}. We can see the Hamiltonian is written in terms of the second-order products of the OAM operators, called the orbital angular position operators~\cite{han2022}, as argued by symmetry analysis. In general, $t_\sigma \neq t_\pi$ and two tangential bands are not degenerated due to the last term in Eq.~(\ref{cubicsp2}), except for some special points. The last term in Eq.~(\ref{cubicsp2}) is absent for previous works~\cite{ko2020,park2022} using a spherical approximation.

\subsection{Mechanism (i)+(ii)}

Now we deal with a triangular lattice with the $s$ and $p$ orbitals where two mechanisms work together. This model has advantages in that, first, it is a more realistic 2D model than a square lattice and, second, it can be shown analytically how the $sp$ hybridization leads to the $r$-$t$ type $p$-orbital texture. We start with the tight-binding Hamiltonian with the ($s$, $p_x$, $p_y$) basis up to the NN hopping.


In a triangular lattice with $sp$ orbitals, both the lattice structure and the $sp$ hybridization mix the $p_x$ and $p_y$ orbitals. We investigate how these two mechanisms work together. The Hamiltonian with the NN hopping is given by
\begin{eqnarray}
\mathcal{H} = \begin{pmatrix} 
H_{ss} & H_{sp_x} & H_{sp_y} \\
H_{sp_x}^* & H_{p_xp_x} & H_{p_xp_y} \\
H_{sp_y}^* & H_{p_yp_x}& H_{p_yp_y}
\end{pmatrix},
\end{eqnarray}
where $H_{p_i p_j}$ is same with Eq.~(\ref{triangularHamiltonian}) where $i,j$ run over $x,y$, and $H_{ss}=E_s^0 + 2t_s \sum_i\cos(\vec{k}\cdot \vec{R}_i)$, $H_{sp_x}= i\gamma_{sp}\sin(\vec{k}\cdot \vec{R}_1)+2i\gamma_{sp}\sin(\vec{k}\cdot \vec{R}_2)+i\gamma_{sp}\sin(\vec{k}\cdot \vec{R}_3),H_{sp_y}= i\sqrt{3}\gamma_{sp}\sin(\vec{k}\cdot \vec{R}_1)-i\sqrt{3}\gamma_{sp}\sin(\vec{k}\cdot \vec{R}_3)$ where $\vec{R}_{1,2,3}$ are defined in Fig.~\ref{Fig:ptriagnular}(a). Then, up to $k^2$ order, the Hamiltonian becomes
\begin{eqnarray} \label{triangularsp}
\mathcal{H}=\begin{pmatrix}
E_s -\frac{3}{2}a^2t_sk^2 & 3ia\gamma_{sp}k_x & 3ia\gamma_{sp}k_y \\
-3ia\gamma_{sp}k_x & E_p+ck^2 + \alpha k_x^2& \alpha k_xk_y \\
-3ia\gamma_{sp}k_y & \alpha k_xk_y & E_p+ck^2 + \alpha k_y^2
\end{pmatrix},
\end{eqnarray}
where $c = -3a^2(t_\sigma+t_\pi)/4, \alpha = -3a^2(t_\sigma-t_\pi)/4.$

Then, we apply the unitary transformation $U_1$ to the Hamiltonian to transform the basis to the ($s$, $p_r$, $p_t$) basis, where $p_r$ is the $p$-orbital parallel to $\vec{k}$, $p_r=\cos(\theta_\vec{k})p_x +\sin(\theta_\vec{k})p_y$ and $p_t$ is the $p$-orbital orthogonal to $\vec{k}$, $p_t = -\sin(\theta_\vec{k})p_x +\cos(\theta_\vec{k})p_y$. Then the transformed Hamiltonian is
\begin{equation}
    U_1\mathcal{H}U_1^\dagger = \begin{pmatrix}
    E_s -\frac{3}{2}a^2t_sk^2  & 3ia\gamma_{sp}k &  0 \\
    -3ia\gamma_{sp}k &E_p+(\alpha+c)k^2 & 0 \\
    0 &0&E_p+ck^2
    \end{pmatrix}.
\end{equation}
As manifested in the transformed Hamiltonian, only the radial orbital is mixed with the $s$ orbital while tangential orbital forms its own band. Due to the hybridization between the $s$ and $p_r$ orbitals, the eigenstate of the previous $p_r$ character band is deformed to $\tilde{p_r}= \cos(\theta_{sp}/2)p_r +i \sin(\theta_{sp}/2)s$ in each $\vec{k}$ direction, where $\theta_{sp} = \tan^{-1}[-3a\gamma_{sp}|k|/(E_s-3a^2t_s/2-E_p-(c+\alpha)k^2)]$. That is, $\tilde{p_r}$ bears imaginary $s$ orbital character. This bears resemblance to the orbital Rashba physics where, due to buckling of lattice, $p_z$ orbital is mixed with $p_r$ orbital in anti-symmetrical way which results in having $\cos(\theta_{\rm ORE})p_r+i\sin(\theta_{\rm ORE})p_z$ as its eigenstates. The $s$ orbital plays the same role as $p_z$ orbital in the buckled lattice in the sense that they lift the degeneracy between the $p_r$ and $p_t$ orbitals. However, whereas $p_r+ip_z$ in the orbital Rashba case forms non-zero OAM, $p_r+is$ has zero OAM. Nevertheless, we have to note that in the $sp$ triangular lattice model, $p_r$-band has $s$ orbital character in the eigenstate and this hybridized states can make some nontrivial physics which we will discuss in Appendix~\ref{Sec(A):pr+is}. From this transformed Hamiltonian, we can infer that this system has the $r$-$t$ type $p$-orbital texture since the energies of the $p_r$ and $p_t$ orbitals are different. We can explicitly show that this is the case by using the L\"owdin downfolding technique. By downfolding the total Hamiltonian into the $p$-orbital subspace and deriving the effective Hamiltonian for $p$ bands in the ($p_x$, $p_y$) basis up $k^2$ order, one obtains
\begin{align} \label{peffhamiltonian}
\mathcal{H}_p &=  \left[3(t_\sigma+t_\pi)-\frac{3a^2(t_\sigma+t_\pi)k^2}{4}\right] \nonumber \\
&\quad +\left[\frac{9a^2\gamma_{sp}^2}{E_p-E_s}- \frac{3a^2}{4}(t_\sigma-t_\pi)\right]\begin{pmatrix}
k_x^2 & k_xk_y \\
k_xk_y & k_y^2
\end{pmatrix}.
\end{align}
This is the same Hamiltonian with Eq.~(\ref{otHamiltonian}) and Fig.~\ref{Fig:2drt} where $\eta = [(t_\sigma-t_\pi)(E_p-E_s)-12\gamma_{sp}^2]/(t_\sigma+t_\pi)(E_p-E_s)$. In this system, both mechanisms drive the system to have the $r$-$t$ type orbital texture. Equation~(\ref{peffhamiltonian}) has two quadratic bands with circular Fermi surface, so it is simple and good platform to model orbital dynamics~\cite{han2022}. 


\section{Advanced models\label{Sec:advanced models}}

In the previous section, we dealt with the simplest models that illustrate the orbital texture generation via the two mechanisms. In this section, we illustrate more general models. First, we change the $sp$ orbitals to the $pd$ orbitals and examine the $p$-orbital texture arising from the $pd$ hybridization. For simplicity, we split the $d$ orbitals into $t_{2g}$ and $e_g$ orbitals and derive the effective Hamiltonians for each group. Then we investigate orbital textures in a hexagonal lattice, a bilayer square lattice, a 3D thin flim with $sp$ orbitals. Accordingly, we find out that additional sublattice degree of freedom plays an important role in forming the orbital texture in a hexagonal lattice and there exists a hidden orbital Rashba texture in a bilayer square lattice. In addition, an anisotropic orbital texture is formed in a 3D thin film model. Lastly, we also derive $p$ and $d$ orbital textures in FCC and BCC structures which are models for realistic 3D materials.

\subsection{Square and cubic lattice with $pd$}

We consider a square lattice with the $pd$ orbitals assuming that the $d$ orbitals are split into $t_{2g}$ and $e_g$ orbitals near the $\Gamma$ point. Then, since these two groups do not hybridize in a square lattice, we can treat them independently and derive the orbital texture for each group. Also, we assume the $p$ orbitals are energetically close to only one of the two groups. 

\subsubsection{$p,t_{2g}$ orbitals in 2D square lattice}

In this subsection, we consider $p$ orbitals and $t_{2g}$ orbitals ($d_{xy}$, $d_{yz}$, $d_{xz}$) on a square lattice. With the mirror symmetry $M_z$, similarly to the case for the $p_z$ orbital above, one can separate these orbitals by the eigenvalues of the mirror operator $M_z$: ($p_x$, $p_y$, $d_{xy}$) for $M_z=1$ and ($p_z$, $d_{xz}$, $d_{yz}$) for $M_z=-1$. These two groups are decoupled and develop orbital texture separately; the first group leads to a $p_x$-$p_y$ orbital texture and the second group leads to a $d_{xz}$-$d_{yz}$ orbital texture. 

\begin{figure}[]
\centering
\includegraphics[width=0.5\textwidth]{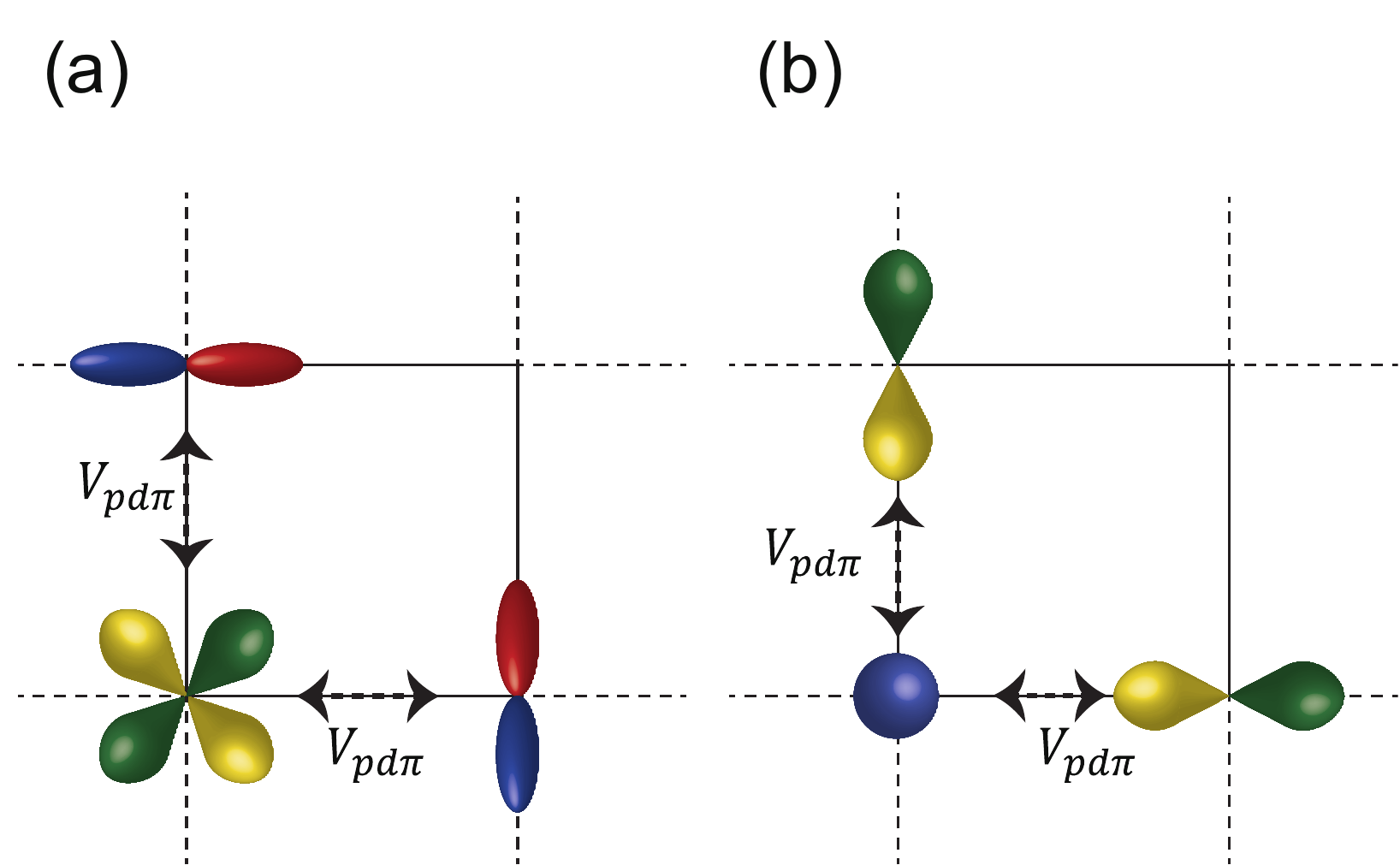}
\caption{\label{Fig:pdhopping}  {Hopping rules for $pd$ orbitals on a square lattice in the $xy$-plane. (a) Hopping rules for $p_x$, $p_y$, $d_{xy}$ orbitals. (b) Hopping rules for $p_z$, $d_{xz}$, $d_{yz}$ orbitals. The blue sphere is the top view of $p_z$ orbital and the other two orbitals (horizontal and vertical yellow-green dumbbells) correspond to the top view of $d_{xz}$, $d_{yz}$ orbitals, respectively.}
}
\end{figure}

First, for the ($p_x$, $p_y$, $d_{xy}$) group, hopping rules are given in Fig.~\ref{Fig:pdhopping}(a). Similar to the $sp$ model, $p_x$ and $p_y$ orbitals are mixed through $d_{xy}$ orbital which leads to an orbital texture. Next, we write down the tight-binding Hamiltonian with NN hoppings in the ($d_{xy}$, $p_x$, $p_y$) basis. 
\begin{equation}
\mathcal{H} = \begin{pmatrix}
H_{d_{xy}d_{xy}} & 2iV_{pd\pi}\sin(ak_y) & 2iV_{pd\pi} \sin(ak_x) \\
-2iV_{pd\pi}\sin(ak_y)& H_{p_xp_x}& 0 \\
-2iV_{pd\pi}\sin(ak_x) & 0 & H_{p_yp_y}
\end{pmatrix},
\end{equation}
where $V_{pd\pi}$ is the $\pi$ hopping integral of the $pd$ orbitals, $H_{d_{xy}d_{xy}}=E_{t_{2g}}^0 +2V_{dd\pi}[\cos(ak_x)+\sin(ak_y)]$, $E_{t_{2g}}$ is the $\Gamma$ point energy of $t_{2g}$ orbital, and $H_{p_xp_x}$ and $H_{p_yp_y}$ are the same as in Eq.~(\ref{sptightbinding}). Now we expand the total Hamiltonian up to $k^2$ order,
\begin{align}
\mathcal{H} &=\begin{pmatrix}
	E_{t_{2g}} & 0&0 \\
	0& E_p& 0 \\
	0 & 0 & E_p
\end{pmatrix}\nonumber \\ 
 &\quad+\begin{pmatrix}
-a^2V_{dd\pi}k^2 & 2iaV_{pd\pi} k_y & 2iaV_{pd\pi} k_x \\
-2iaV_{pd\pi} k_y& -a^2(t_\sigma k_x^2+t_\pi k_y^2)& 0 \\
-2iaV_{pd\pi} k_x & 0 & -a^2(t_\pi k_x^2 +t_\sigma k_y^2)
\end{pmatrix}.
\end{align}
Similar with the $sp$ model, the $pd$ hybridization ($V_{pd\pi}$) opens the gap at $k_x = \pm k_y$ points and the difference of the energies of $p$ and $d$ orbitals determines the type of orbital texture.\footnote{There exists a difference between the $sp$ and $pd$ models. The $sp$ hybridization renormalizes $t_\sigma$ while the $pd$ hybridization renormalizes $t_\pi$ hopping. This stems from different hopping rules between $sp$ and $pd$ orbitals. However, near degeneracy points, they play the same roles.} This can be explicitly checked by the L\"owdin downfolding technique. We downfold the total Hamiltonian to the $p$-character bands and obtain
\begin{equation} \label{pdHamiltonian}
    \mathcal{H}_p =\mathcal{H}_p^{\rm NN} +\frac{4a^2V_{pd\pi}^2}{E_p-E_{t_{2g}}}\begin{pmatrix}
    k_y^2 & k_xk_y\\
    k_xk_y & k_x^2
    \end{pmatrix}.
\end{equation}
By the same way as the $sp$ case, the off-diagonal components in Eq.~(\ref{pdHamiltonian}) separate the two bands and form the orbital texture in Fig.~\ref{Fig:rashbadressel}.



\begin{figure}
\centering
\includegraphics[width=0.5\textwidth]{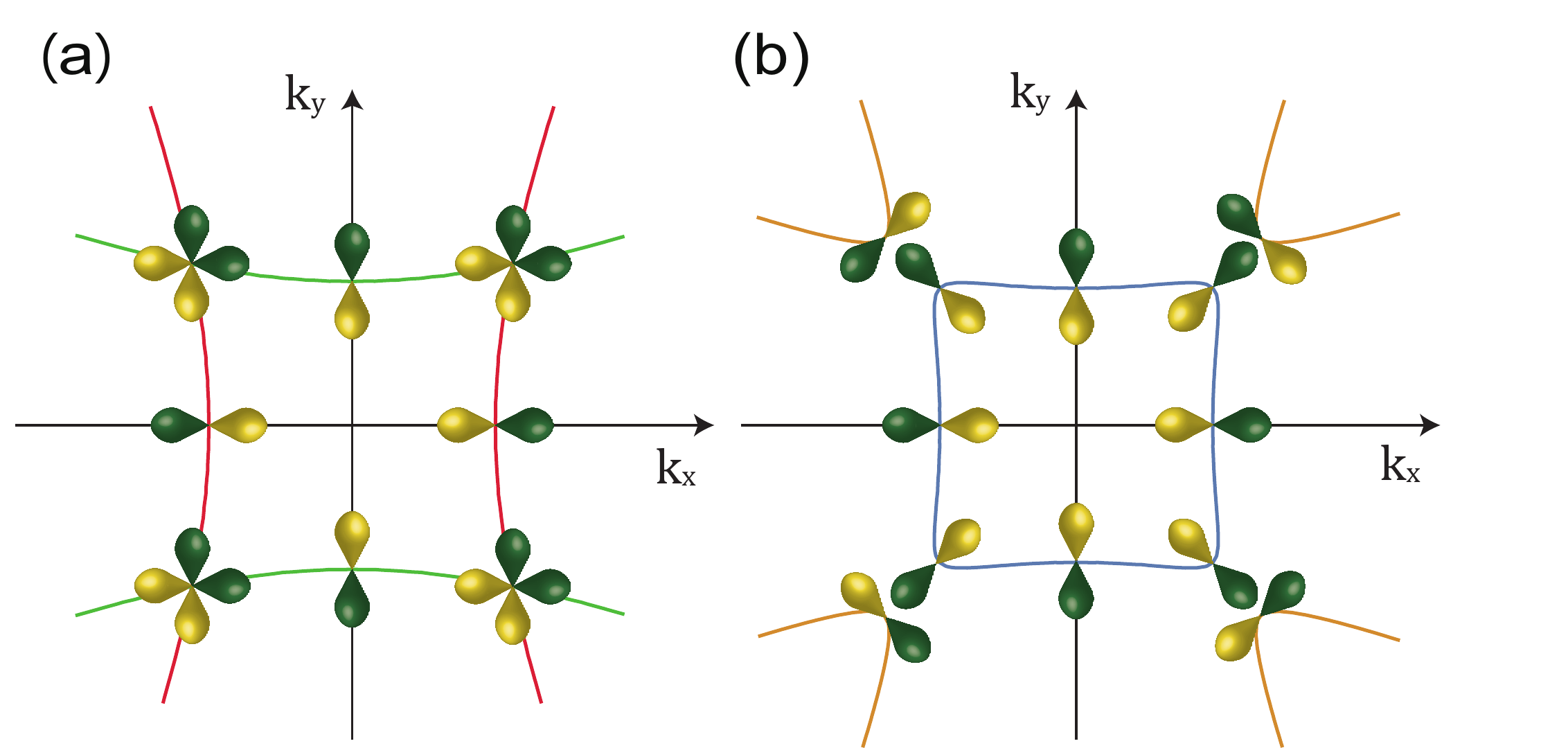}
\caption{\label{Fig:t2gtexture}  {$t_{2g}$ orbital character in a square lattice. We plot top view of $d_{xz}$ and $d_{yz}$ orbitals. (a) The Fermi surface and orbital characters up to NN hopping. $d_{xz}$ and $d_{yz}$ orbitals are degenerate for the directions $k_x=\pm k_y$. There is no orbital texture. (b) Orbital texture due to NNN hopping. The $r$-$t$ type texture is formed for $E_p-E_d<0$ (shown here) while the \textit{Dresselhaus} type texture is formed for $E_p-E_d>0$ (not shown).}
}
\end{figure}

Second, we deal with the other group, ($p_z$, $d_{yz},$ $d_{zx}$), for which hopping rules are given in Fig.~\ref{Fig:pdhopping}(b). We downfold the total Hamiltonian into the $d$-orbital bands and obtain the effective Hamiltonian in the ($d_{yz},$ $d_{zx}$) basis, which is given by
\begin{align}
\mathcal{H}_{t_{2g}} &= E_{t_{2g}}-a^2 \begin{pmatrix}
V_{dd\delta}k_x^2 + V_{dd\pi}k_y^2 & 0 \\
0& V_{dd\pi}k_x^2 + V_{dd\delta}k_y^2
\end{pmatrix} \nonumber \\
 &\quad +\frac{4a^2V_{pd\pi}^2}{E_{t_{2g}}-E_p} \begin{pmatrix}
k_y^2 & k_xk_y \\
k_xk_y & k_x^2
\end{pmatrix}.
\end{align}
This is similar to the result for the square lattice model with the $sp$ orbitals [Eq.~(\ref{squareporbital})], if we map orbitals as follows: $d_{xz}\rightarrow p_x,d_{yz}\rightarrow p_y$. Similar to the $sp$ case, the gaps are opened by the $pd$ hybridization and the difference of $E_p$ and $E_d$ determines whether the texture is the $r$-$t$ type or the \textit{Dresselhaus} type (Fig.~\ref{Fig:t2gtexture}).

\subsubsection{$p,t_{2g}$ orbitals in 3D cubic lattice}

We generalize the 2D square lattice to a 3D cubic lattice structure with $p$ and $t_{2g}$ orbitals. In this case, all $p$-$t_{2g}$ orbitals are hybridized and we derive the total Hamiltonian. For simplicity, we show only the effective Hamiltonian after the downfolding procedure and demonstrate that its physical meanings are similar to the square-lattice case. 

First, we downfold the total Hamiltonian to the $p$-orbital subspaces. The effective Hamiltonian in the ($p_x$, $p_y$, $p_z$)  basis is given by
\begin{align} \label{pbypd}
\mathcal{H}_p &= E_p - a^2 t_\pi k^2 -a^2(t_\sigma-t_\pi)\begin{pmatrix}
k_x^2  & 0 & 0 \\
0 &  k_y^2& 0 \\
0& 0 & k_z^2 \end{pmatrix}\nonumber \\ &\quad + \frac{4a^2V_{pd\pi}^2}{E_p-E_{t_{2g}}}\begin{pmatrix}
k_y^2+k_z^2 & k_xk_y & k_xk_z\\
k_xk_y & k_x^2+k_z^2 & k_yk_z\\
k_xk_z & k_yk_z & k_x^2+k_y^2 
\end{pmatrix}.
\end{align}
This Hamiltonian can also be decomposed into the second-order products of the OAM operators as,
\begin{align}
\mathcal{H}_p &= E_p -a^2t_\sigma k^2  -\frac{2a^2V_{pd\pi}^2}{E_p-E_{t_{2g}}}\sum_{i,j(\neq i)}k_ik_j \{L_i,L_j\}
 \nonumber \\
&\quad +\left[a^2(t_\sigma-t_\pi)+\frac{4a^2V_{pd\pi}^2}{E_p-E_{t_{2g}}}\right]\sum_ik_i^2L_i^2,
\end{align}
where $\{L_i,L_j\}= L_iL_j +L_jL_i$.

Similarly, we downfold the total Hamiltonian to the $d$-orbital subspace and obtain the effective Hamiltonian in the ($d_{yz}$, $d_{zx}$, $d_{xy}$) basis:
\begin{align} \label{t2gbypd}
\mathcal{H}_{t_{2g}} &=E_{t_{2g}}-a^2V_{dd\pi}k^2  +a^2(V_{dd\pi}-V_{dd\delta}) \begin{pmatrix}
k_x^2 & 0 & 0 \\
0& k_y^2 & 0 \\
0& 0 & k_z^2
\end{pmatrix} \nonumber \\
&\quad+ \frac{4a^2V_{pd\pi}^2}{E_{t_{2g}}-E_p}\begin{pmatrix}
k_y^2+k_z^2& k_xk_y & k_xk_z\\
k_xk_y & k_x^2+k_z^2  & k_yk_z\\
 k_xk_z& k_yk_z& k_x^2+k_y^2
\end{pmatrix}.
\end{align}
Similar to the $p$-orbital case, this Hamiltonian can be decomposed by the OAM operators of $t_{2g}$-orbitals. When one projects $L_i$ operators of $d$-orbitals onto $t_{2g}$ orbitals, they have the same mathematical structure to those of the $p$ orbitals by mapping $d_{yz} \rightarrow p_x$, $d_{zx} \rightarrow p_y$ and $d_{xy} \rightarrow p_z$~\cite{kim2008}. We thus use the same $L_i$ symbol to the $t_{2g}$-orbital operators\footnote{Since $\vec{L}$ defined in this way satisfies a modified angular momentum commutation relation (with an additional negative sign), an extra care is needed when one deals with dynamic phenomena or when the time-reversal symmetry is broken.}. Then Hamiltonian in terms of $L_i$ is written as
\begin{align}
\mathcal{H}_{t_{2g}} &= E_{t_{2g}} -a^2V_{dd\delta}k^2 -\frac{2a^2V_{pd\pi}^2}{E_{t_{2g}}-E_p}\sum_{i,j(\neq i)}k_ik_j \{L_i,L_j\}
\nonumber\\
&\quad+a^2\left[(V_{dd\delta}-V_{dd\pi})+\frac{4V_{pd\pi}^2}{E_{t_{2g}}-E_p}\right]\sum_i k_i^2L_i^2.
\end{align}

\subsubsection{$p,e_g$ orbitals in 2D square lattice}

In this subsection, we discuss the orbital texture formed by the $p$-$e_g$ hybridization on a square lattice. We consider ($p_x$, $p_y$, $d_{z^2}$, $d_{x^2-y^2}$) orbitals on a square lattice. We assume that the two $e_g$ orbitals are degenerate at the $\Gamma$ point. For this case, the hybridizations between ($p_x,p_y$)-$e_g$ orbitals lead to both a $p$-orbital texture and an $e_g$-orbital texture. 

\begin{figure}[]
\centering
\includegraphics[width=0.3\textwidth]{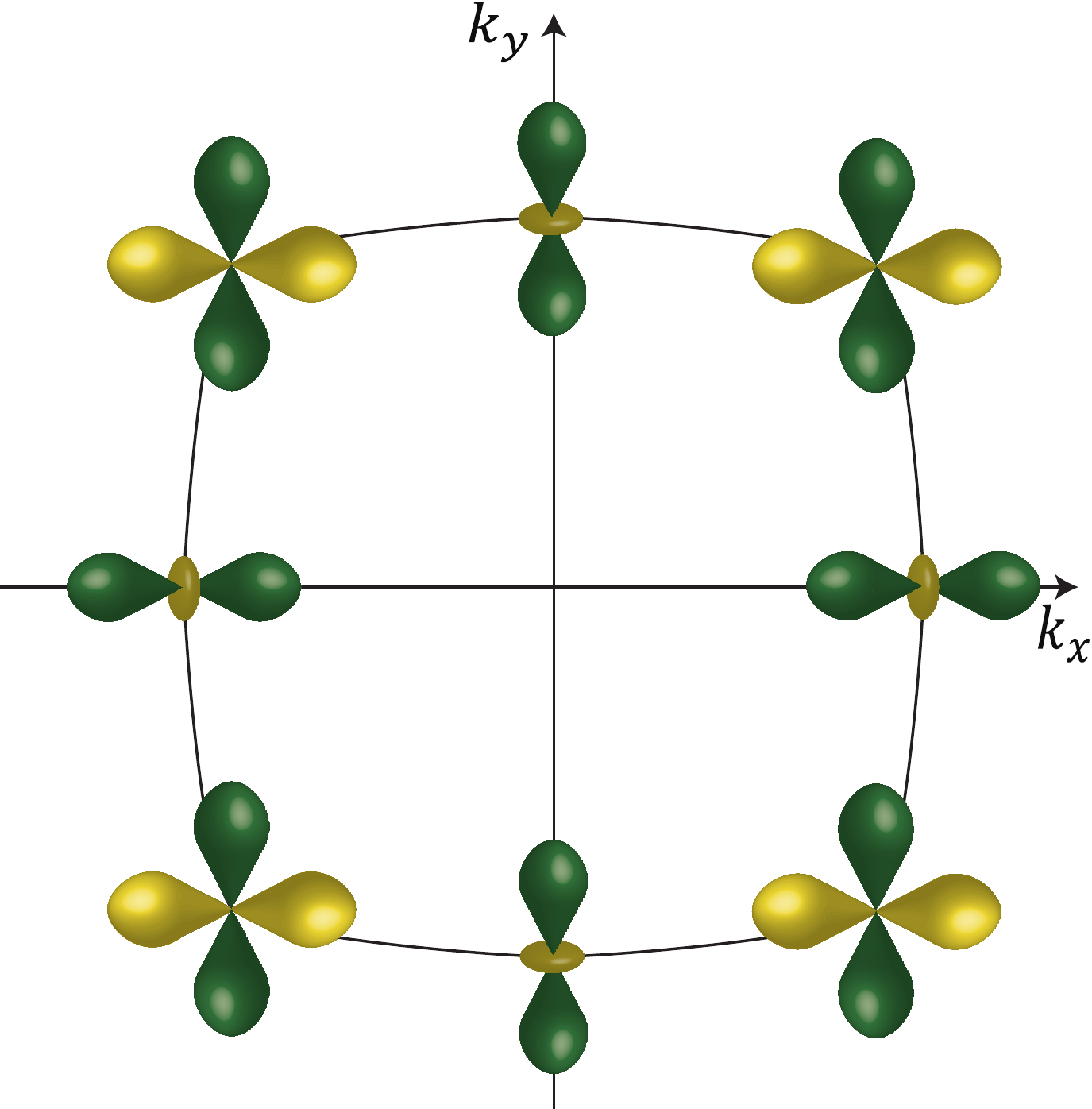}
\caption{\label{Fig:egtexture} $e_g$-orbital texture generated by $\mathcal{H}_{eg}$ in Eq.~(\ref{egtexture}). Because of complexity, we only plot the inner band of the Hamiltonian.}
\end{figure}

Let us look at the $e_g$-orbital texture driven by mechanism~(i) first. Note that $e_g$ orbitals hybridize with each other through NN hoppings so that mechanism (i) does not necessarily require NNN hoppings. The $e_g$-orbital Hamiltonian in the ($d_{x^2-y^2}$, $d_z^2$) basis up to $k^2$ order is given by
\begin{align} \label{egtexture}
\mathcal{H}_{e_g}&= E_{e_g}-\frac{a^2(V_{dd\delta}+V_{dd\pi})k^2}{2} \nonumber \\
&\quad +\frac{a^2(V_{dd\delta}-V_{dd\pi})}{4}\begin{pmatrix}
k^2 & -\sqrt{3}(k_x^2-k_y^2) \\
-\sqrt{3}(k_x^2-k_y^2) & -k^2
\end{pmatrix},
\end{align}
where $E_{e_g}$ is the $\Gamma$ point energy of the $e_g$ orbitals. Their eigenstates are linear combinations of $e_g$ orbitals which are plotted in Fig.~\ref{Fig:egtexture}. For $k_y=0$, for example, the eigenstate of the inner band is $(\sqrt{3}d_{x^2-y^2}-d_{z^2})/2$ which is $d_{x^2}$ orbital.

Next, we consider the $pe_g$ hybridization [mechanism (ii)] and apply the L\"owdin downfolding to the $e_g$-orbital subspace. The resulting Hamiltonian is
\begin{equation} 
\Delta \mathcal{H}_{e_g} = 2\beta k^2+ \beta\begin{pmatrix}
 k^2 & -\sqrt{3}(k_x^2-k_y^2) \\
-\sqrt{3}(k_x^2-k_y^2) & -k^2 
\end{pmatrix},
\end{equation}
where $\Delta \mathcal{H}_{e_g}$ is the additional $e_g$-orbital Hamiltonian due to the $pe_g$ hybridization and $\beta= a^2V_{pd\sigma}^2/(E_{e_g}-E_p)$. Interestingly, $\Delta H_{e_g} $ commutes with $H_{e_g}$, which means that the $e_g$-orbital texture driven by $p$ orbitals is of the same form as the $e_g$ orbital's own texture [Eq.~(\ref{egtexture})]. In other words, the mechanism (ii) simply renormalizes the strength of the orbital texture in Eq.~(\ref{egtexture}).

For the $p$-orbital texture driven by the hybridization with $e_g$ orbitals, we downfold the total Hamiltonian to the $p$-orbital subspace and obtain
\begin{equation}
\mathcal{H}_{p}= \mathcal{H}_p^{\rm NN} +
\frac{2a^2 V_{pd\sigma}^2}{E_p-E_{e_g}}\begin{pmatrix}
2k_x^2 & -k_xk_y \\
-k_xk_y & 2k_y^2
\end{pmatrix}.
\end{equation}
Thus the gaps are opened by the $pd$ hybridization ($V_{pd\sigma}$) and, similar with the previous cases, the difference of $E_p$ and $E_{e_g}$ determines the type of the orbital texture (Fig.~\ref{Fig:rashbadressel}).

\subsubsection{$pe_g$ orbitals in 3D cubic lattice}

We now generalize the 2D square lattice to the 3D cubic lattice with $pe_g$ orbitals. By using the same formalism as above, the $e_g$ orbital-texture Hamiltonian driven by the mechanism (i) turns out to be
\begin{align} \label{3degtexture}
\mathcal{H}_{e_g,0} &= E_{e_g}-\frac{a^2(V_{dd\delta}+V_{dd\pi})k^2}{2} \nonumber \\
 &\quad +\frac{a^2(V_{dd\delta}-V_{dd\pi})}{4} \begin{pmatrix}
k^2-3k_z^2 & -\sqrt{3}(k_x^2-k_y^2)\\
-\sqrt{3}(k_x^2-k_y^2) &-k^2+3k_z^2
\end{pmatrix}.
\end{align}

For mechanism (ii), the additional $e_g$-orbital Hamiltonian due to the $pe_g$ hybridization is given by
\begin{equation} \label{egbypd}
\Delta \mathcal{H}_{e_g}=2\beta k^2 +  \beta\begin{pmatrix}
k^2-3k_z^2& -\sqrt{3}(k_x^2-k_y^2)\\
-\sqrt{3}(k_x^2-k_y^2) &-k^2+3k_z^2
\end{pmatrix}.
\end{equation}
For the $p$-orbital subspace, the $p$-orbital texture Hamiltonian driven by the $pe_g$ hybridization becomes, 
\begin{align} 
\mathcal{H}_p&=E_p - a^2t_\pi k^2 -a^2(t_\sigma-t_\pi)\begin{pmatrix}
k_x^2  & 0 & 0 \\
0 &  k_y^2& 0 \\
0& 0 & k_z^2 \end{pmatrix}\nonumber \\
&\quad + \frac{2a^2 V_{pd\sigma}^2}{E_p-E_{e_g}}\begin{pmatrix}
2k_x^2 & -k_xk_y & -k_xk_z\\
-k_xk_y & 2k_y^2 & -k_yk_z\\
-k_xk_z & -k_yk_z & 2k_z^2
\end{pmatrix},\label{peg cubic}
\end{align}
in the ($p_x$, $p_y$, $p_z$) basis. Note that the effective Hamiltonian resulting from the $pe_g$ hybridization and its physical meanings are same as the 2D square lattice case. 

As a remark, Eq.~(\ref{peg cubic}) can be expressed in terms of the OAM operator as
\begin{align}
\mathcal{H}_p &= E_p -\left(a^2 t_\sigma -\frac{4a^2V_{pd\sigma}}{E_p-E_{e_g}}\right) k^2 +\frac{a^2V_{pd\sigma}}{E_p-E_{e_g}}\sum_{i,j\neq i}k_ik_j \{L_i,L_j\} \nonumber \\
 &\quad + \left[a^2(t_\sigma-t_\pi)-\frac{4a^2V_{pd\sigma}}{E_p-E_{e_g}}\right] \sum_i k_i^2L_i^2 .
\end{align}

\subsection{Hexagonal lattice with $sp$}
In this subsection, we derive the effective Hamiltonian for the hexagonal lattice with $sp$ orbitals near the $\Gamma$ and K points. This system is complex in that it has additional sublattice degree of freedom.
We label the two sublattices as A and B, respectively. This is a generalized version of graphene in the sense that it has the $p_x$- and $p_y$-orbital characters while the pristine graphene has the $p_z$-orbital character near the Fermi energy.
This system is intensively studied in the field of the orbital-filtering effect~\cite{zhang2021,sun2021}. $p$-doped graphane and BiH~\cite{tokatly2010,hao2022} are the corresponding real materials. By using our formalism, we derive the effective $p$-band Hamiltonian, which is consistent with the previous study~\cite{wu2008}.


\subsubsection{Hamiltonian near the $\Gamma$ point.}

At the $\Gamma$ point, the sublattice degree of freedom can be easily handled because the $p$-orbital sublattice-symmetric states ($p\otimes  (\ket{A}+\ket{B})/\sqrt{2}$) and the $p$-orbital sublattice-anti-symmetric states ($p\otimes (\ket{A}-\ket{B})/\sqrt{2}$) are gapped by $\propto t_\sigma+t_\pi$, which is on the order of eV. Therefore, these symmetric and anti-symmetric states are separated well and one can safely apply the L\"owdin-downfolding technique to each state. As a result, the effective Hamiltonian of the sublattice-anti-symmetric $p$-orbital state in $(p_x\otimes  (\ket{A}-\ket{B})/\sqrt{2},p_y\otimes  (\ket{A}-\ket{B})/\sqrt{2})$ basis is given,
\begin{align}
\mathcal{H}_{p} &=\tilde{E_p} +\frac{3a^2(t_\sigma+3t_\pi)k^2}{16}  \nonumber \\ 
&\quad +  \left[ \frac{3a^2}{8}(t_\sigma-t_\pi) + \frac{9a^2\gamma_{sp}^2}{4(\tilde{E_p}-\tilde{E_s})}\right] 
\begin{pmatrix}
k_x^2& k_xk_y \\
k_xk_y & k_y^2
\end{pmatrix},
\end{align}
where $\tilde{E_p} = E_p^0-3(t_\sigma+t_\pi)/2$ and $\tilde{E_s}=E_s^0-3t_s$ are the $\Gamma$-point energy of the sublattice-anti-symmetric $p$ orbital and that of the sublattice-anti-symmetric $s$ orbital, respectively. This corresponds to the orbital texture in Fig~\ref{Fig:2drt} and similar to the triangular lattice with $sp$ model, both lattice structure and $sp$ hybridization drive system to have an $r$-$t$ type orbital texture. Similar results can be derived for the sublattice-symmetric states. That is, a hexagonal lattice exhibits orbital textures for sublattice-symmetric and sublattice-anti-symmetric states respectively, which is consistent with previous report~\cite{wu2008}. As a remark, by the same reason of the triangular lattice, a hexagonal lattice bears the orbital texture even without the $sp$ hybridization and there is no warping terms up to $k^2$ order due to $\pi/3$-rotation symmetry.

\subsubsection{Hamiltonian near the K point.}

At the K point, there are two degenerate states: $p_x-ip_y$ orbital at the A sublattice [$1/\sqrt{2}(p_x-ip_y)\otimes \ket{A}$] and $p_x+ip_y$ orbital at the B sublattice [$1/\sqrt{2}(p_x+ip_y)\otimes\ket{B}$]. We investigate the effective Hamiltonian formed by these two degenerated states near the K point. For the simplicity, we set $\gamma_{sp}$ to be zero. Then, effective Hamiltonian is given,
\begin{equation}
\mathcal{H}_{p}=\frac{3a(t_\sigma-t_\pi)}{4}\left[\left(\frac{q_x}{2}-\frac{\sqrt{3}q_y}{2}\right)\sigma_x+\left(\frac{\sqrt{3}q_x}{2}+\frac{q_y}{2}\right)\sigma_y \right],
\end{equation}
where $\vec{q}=\vec{k}-\vec{k}_{\rm K}$ for the K point momentum $\vec{k}_{\rm K}$, and $\sigma_x$ and $\sigma_y$ are the pseudospin Pauli matrices in the $[(p_x-ip_y)/\sqrt{2}\otimes \ket{A}$, $(p_x+ip_y)/\sqrt{2}\otimes \ket{B}]$ basis. The effective Hamiltonian near $K'$ point can be derived by applying the time-reversal operation.

A few remarks are in order. First, the generalized version of graphene also has the Dirac states near the K point\footnote{Energies of two states are given by $E_{\pm}= \pm 3a(t_\sigma-t_\pi)|q|/4$.}. However, the pseudospin basis is different from the pristine graphene. While the pristine graphene has $p_z$ orbital with the two sublattice states as the pseudospin basis, our model has the A sublattice with the $L_z=-1$ eigenvalue and the B sublattice with the $L_z=1$ eigenvalue as its basis~\cite{wu2008}. The pseudospin shares a similar structure with the spin Pauli matrix under the time reversal operator. But unlike the spin Pauli matrix, which describes spin angular momentum in different directions, the pseudospin describes only the OAM in the direction of the $L_z$. Second, the degenerated states in the K point have the quantum number $+1$ and $-1$ for the $L_z$ operator, so the effect of spin-orbit coupling would be greater than the pristine graphene which has only the $p_z$ character. Furthermore, the spin-orbit coupling term comes into $\lambda S_z \sigma_z$ in our pseudospin basis where $\lambda$ is the strength of spin-orbit coupling, so it is easy to model the effect near the K point using our Hamiltonian. An inversion symmetry breaking term from an on-site potential difference between the A and B sublattices can be easily introduced by adding $\delta \sigma_z$ where $\delta$ is the on-site energy difference between the A and B sublattices.

\subsection{Bilayer square lattice with $sp$}

\begin{figure}
\centering
\includegraphics[width=0.4\textwidth]{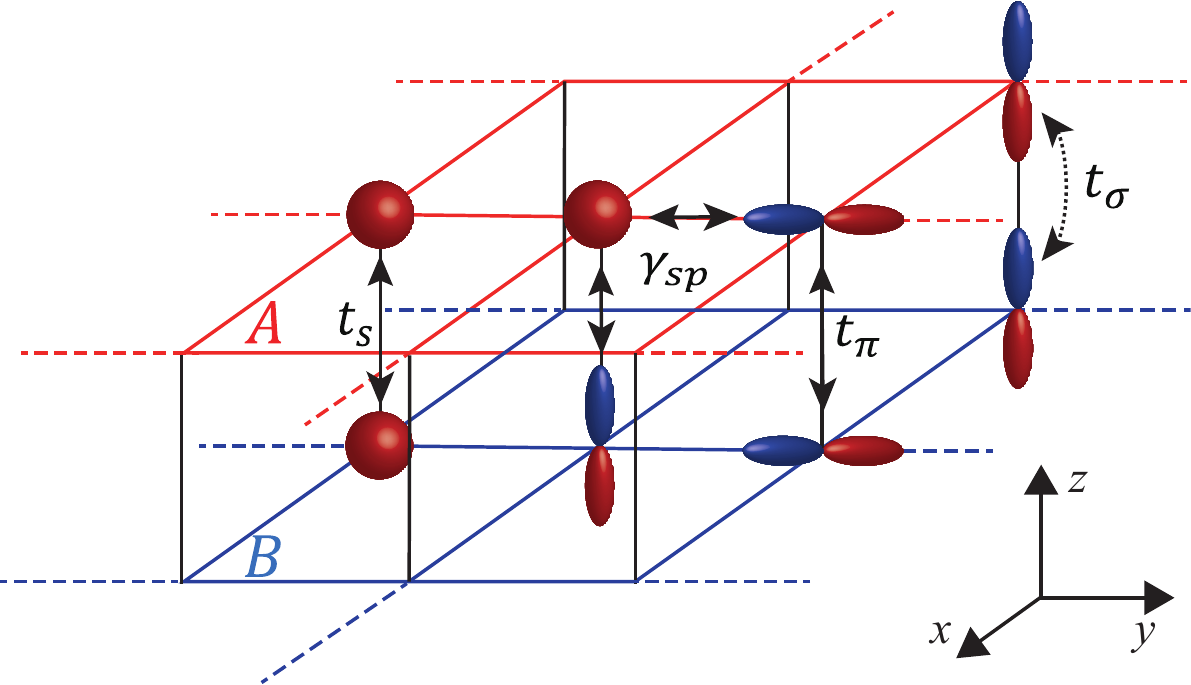}
\caption{\label{Fig:bilayer} Schematic of a bilayer square lattice and the according hopping rules. Red layer denotes the A layer and blue layer denotes the B layer.}
\end{figure}

In this case, we derive an effective Hamiltonian in a bilayer square lattice with the $sp$ orbitals (Fig.~\ref{Fig:bilayer}). For this case, we keep the $p_z$ orbital under our consideration. We assume that the on-site energies of the in-plane $p$ orbitals ($p_x$, $p_y$) and the out-of-plane $p$ orbital ($p_z$) are different ($E_{p_x,p_y}=E_{p_1} \neq E_{p_z}=E_{p_2}$). We label the upper layer as A layer and the lower layer as B layer. The hopping rules are given in Fig.~\ref{Fig:bilayer}. This is similar to the square lattice with the $sp$-orbital model but different in that the $p_z$ ($s$) orbital in A layer is mixed with the $s$ ($p_z$) orbital in B layer. This hybridization makes different orbital textures from that of the square $sp$ model. Furthermore, we show below that the antisymmetric hopping between $p_z$ and $s$ orbitals generates a \textit{hidden orbital Rashba} texture.

We construct the $8\times8$ bilayer Hamiltonian $\mathcal{H}$ based on the hopping rules in Fig.~\ref{Fig:bilayer}:
\begin{equation}
\mathcal{H}= \begin{matrix}
& \begin{matrix}
\ket{A}& \ket{B}
\end{matrix} \\
\begin{matrix}
\bra{A}\\
\bra{B}
\end{matrix} &\begin{pmatrix}
H_{A} & H_{AB} \\
H_{BA} & H_{B}
\end{pmatrix}
\end{matrix},
\end{equation}
where $H_{A}$ and $H_{B}$ denote the intralayer Hamiltonians while $H_{AB}$ and $H_{BA}$ denote the interlayer coupling Hamiltonians. First, the intralayer Hamiltonians ($H_{A},H_{B}$) are the same as that of the square $sp$ model. Up to $k^2$ order, the intralayer Hamiltonians in the ($s$, $p_x$, $p_y$, $p_z$) basis are
\begin{equation}
H_{A}=H_{B}=\begin{pmatrix}
H_{ss} & 2ia\gamma_{sp}k_x & 2ia\gamma_{sp}k_y & 0 \\
-2ia\gamma_{sp}k_x &H_{p_xp_x} & 0& 0 \\
-2ia\gamma_{sp}k_y & 0 & H_{p_yp_y}& 0 \\
0&0&0&H_{p_zp_z}
\end{pmatrix},
\end{equation}
where $H_{ss}=E_s -a^2t_sk^2$, $H_{p_xp_x/p_yp_y}= E_{p_1}-a^2(t_{\sigma/\pi} k_x^2+t_{\pi/\sigma} k_y^2)$, and $H_{p_zp_z}=E_{p_2}-a^2t_\pi k^2$. Next, for the interlayer coupling, the $p_z$ ($s$) orbital in the A layer is mixed with the $s$ ($p_z$) orbital in the B layer. In addition, the $s$ and $p$ orbitals are mixed with the same orbitals in the other layer: the $t_s$ hopping between the $s$ orbitals and the $t_\pi$ hopping between the in-plane $p$-orbitals, and the $t_\sigma$ hopping between the $p_z$ orbitals along the out-of-plane direction. The matrix representation of the hopping rule is given by
\begin{equation}
H_{AB}= H_{BA}^\dagger=\begin{pmatrix}
t_s & 0 & 0 &-a\gamma_{sp} \\
0& t_\pi & 0 &0 \\
0& 0  & t_\pi & 0 \\
a\gamma_{sp} &0&0&t_\sigma
\end{pmatrix}.
\end{equation}

Similar to the $sp$ square lattice model, the orbitals are divided by the eigenvalues of the $M_z$ operator. Considering the additional layer degree of freedom, the $\sigma$-bonding $s$, the $\pi$-bonding $p_x,p_y$ orbitals (symmetric superposition in the two layers) and the $\sigma$-bonding $p_z$ orbital (antisymmetric superposition in the two layers) have $M_z=+1$ eigenvalue while the $\sigma^*$-antibonding $s$, $\pi^*$-antibonding $p_x,p_y$ orbitals (antisymmetric superposition in the two layers) and $\sigma^*$-antibonding $p_z$ (symmetric superposition in the two layers) orbitals have $M_z=-1$ eigenvalue. These two groups do not hybridize with each other. More explicitly, we perform the unitary transformation $U_1$ that block-diagonalizes the $M_z=\pm1$ blocks.
\begin{equation} \label{bilayerhamiltonian}
U_1 \mathcal{H} U_1^\dagger = \begin{pmatrix}
H_{+1} & 0 \\
0& H_{-1}
\end{pmatrix},
\end{equation}
where $H_{\pm 1}$ is the reduced Hamiltonian for the subspace for $M_z=\pm1$, whose explicit expression is given by
\begin{equation}
H_{\pm 1} =  \begin{pmatrix}
H_{ss}\pm t_s & 2ia\gamma_{sp}k_x & 2ia\gamma_{sp}k_y & \mp a\gamma_{sp} \\
-2ia\gamma_{sp}k_x & H_{p_xp_x}\pm t_\pi & 0 &0 \\
-2ia\gamma_{sp}k_y & 0 & H_{p_yp_y} \pm t_\pi & 0 \\
-a\gamma_{sp} & 0 & 0& H_{p_zp_z}\mp t_\sigma
\end{pmatrix}.
\end{equation}

We focus on the $H_{+1}$ groups since the same analysis applies to $H_{-1}$. We downfold the $H_{+1}$ Hamiltonian to the $p$-character band using the L\"owdin downfolding technique and obtain
\begin{align} \label{hiddenore}
H_{+1,p} &= \begin{pmatrix}
 H_{p_xp_x}+t_\pi& 0& 0 \\
0&H_{p_yp_y}+t_\pi & 0 \\
 0&0& H_{p_zp_z}-t_\sigma
\end{pmatrix}\nonumber  \\  
&\quad +\alpha \begin{pmatrix}
 k_x^2 &  k_xk_y & 0 \\
 k_xk_y&  k_y^2 & 0 \\
 0&0& \delta
\end{pmatrix} + \beta (k_y L_x - k_x L_y),
\end{align}
where $\alpha=4a^2\gamma_{sp}^2/(E_{p_1}+t_\pi-E_s-t_s)$, $\beta=a^2\gamma_{sp}^2/(E_{p_1}+t_\pi-E_s-t_s)+a\gamma_{sp}^2/(E_{p_2}+t_\sigma-E_s-t_s),$ and $\delta=a\gamma_{sp}^2/(E_{p_2}-t_\sigma-E_s-t_s)$. Terms proportional to $\alpha$ in Eq.~(\ref{hiddenore}) corresponds to the orbital texture driven by the $sp$ hybridization as in the square lattice model.

 The new term proportional to $\beta$ corresponds to the \textit{orbital Rashba effect} $\propto \vec{L}\cdot(\vec{k} \times\vhat{z})$~\cite{park2011}. The $\pi$-bonding ($p_x,p_y$) states are mixed with $\sigma$-bonding $p_z$ states through the $s$ orbital and form the orbital Rashba states. Actually these are not a genuine orbital Rashba texture since $\vec{L}$ is not a genuine OAM operator. The $\pi$-bonding ($p_x,p_y$) states are of the form $p_{x,y}\otimes \begin{bmatrix}
\frac{1}{\sqrt{2}}\\
\frac{1}{\sqrt{2}}
\end{bmatrix} $ while the $\sigma$-bonding $p_z$ state is $p_z \otimes \begin{bmatrix}
\frac{1}{\sqrt{2}}\\
\frac{-1}{\sqrt{2}}
\end{bmatrix}$, and since the eigenstates of $L_x,L_y$ operators in Eq.~(\ref{hiddenore}) are superpositions of $p_{x,y}$ and $p_z$ orbitals, this leads to layer-opposite OAM structures
, resulting in vanishing total OAM in equilibrium for eigenstate $\vec{k}$; when the eigenstate is projected onto the upper layer or lower layer, the individual layer have the OAM expectation values of the opposite sign thus compensating each other. In this sense, this is a hidden orbital Rashba texture. However, one can induce a non-compensated orbital Rashba states by applying a vertical voltage which makes the cancellation between the layers incomplete so that the total orbital Rashba interaction becomes nonzero. In addition, when the two layers are not equivalent due to, for instance, different material parameters or a work function difference, the hidden orbital Rashba state is naturally converted to a nonvanishing orbital Rashba state and thus may explain the previous experimental observations~\cite{tsai2018,park2018} in the presence of spin-orbit coupling. This is a different mechanism from the previously reported mechanism for the orbital Rashba effect~\cite{park2013,sunko2017}. 

\subsection{Thin films in 3D} 

\begin{figure}
	\centering
	\includegraphics[width=0.4\textwidth]{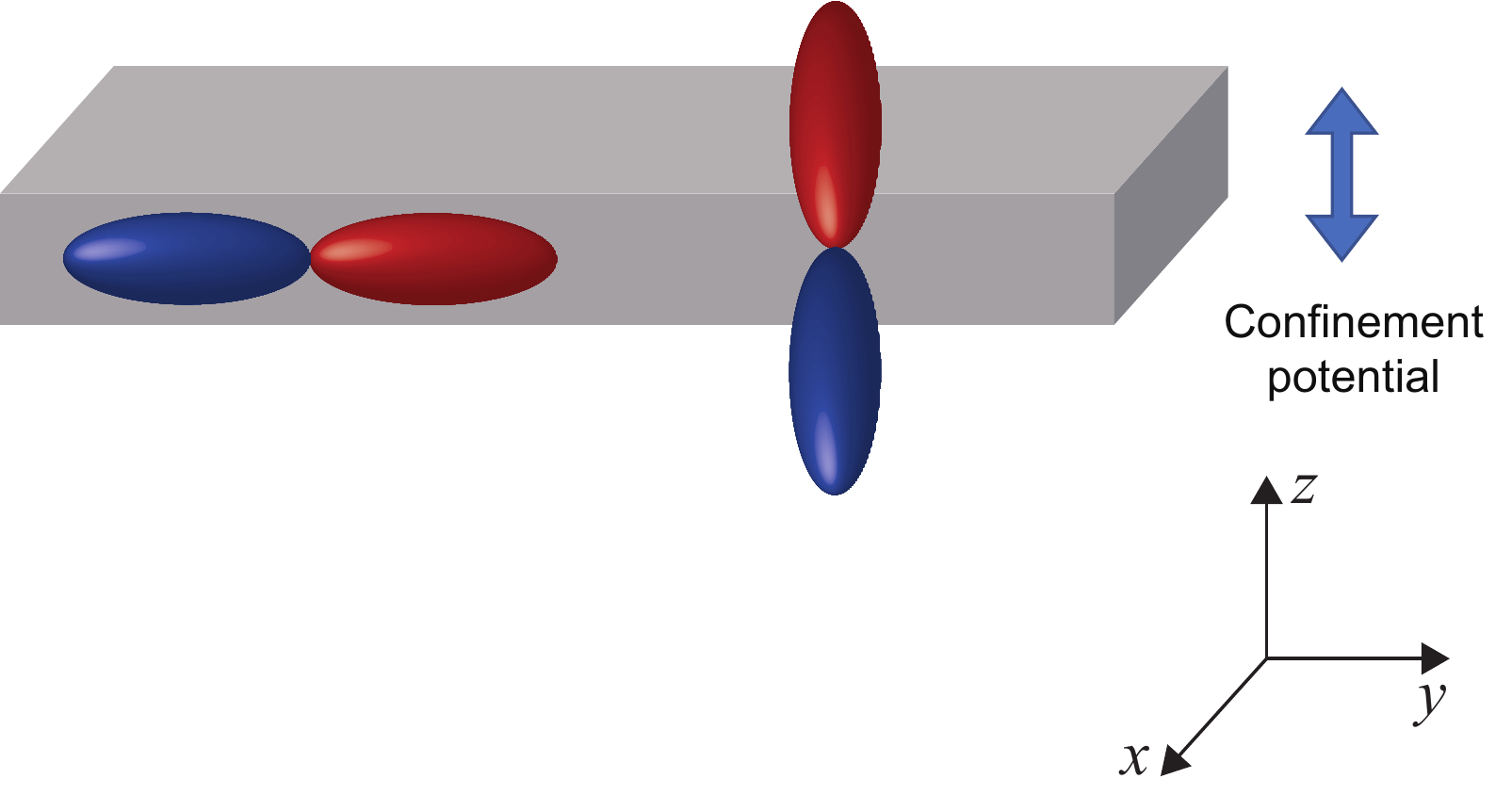}
	\caption{\label{Fig:thinfilm} Schematic of the 3D thin film model. $E_{p_x,p_y}$ and $E_{p_z}$ can be different since there exists the confinement potential along the $z$ axis.
	}
\end{figure}

In 3D cubic lattices demonstrated above, we have considered isotropic materials where the on-site energies of $p_x$, $p_y$, and $p_z$ orbitals are the same. However, in thin films with a finite thickness along the $z$ direction, the onsite energy of $p_z$ orbital is in general different from that of the in-plane orbitals (Fig.~\ref{Fig:thinfilm}). In this subsection, we derive the orbital texture in a thin film model in 3D (considering $k_z$) which is the continuum generalization of the bilayer Hamiltonian in the previous subsection. We use the same hopping rules given by the cubic lattice. 

After performing L\"owdin-downfolding to the $sp$ hybridization term, the effective $p$-orbital Hamiltonian is given by
\begin{align} \label{thinfilmHamiltonian}
\mathcal{H}_p &= \begin{pmatrix}
E_{p_1} &0&0\\
0& E_{p_1} & 0 \\
0&0&E_{p_2}
\end{pmatrix} - a^2 t_\pi k^2 -a^2(t_\sigma-t_\pi)\begin{pmatrix}
 k_x^2  & 0 & 0 \\
0 &   k_y^2& 0 \\
0& 0 & k_z^2
\end{pmatrix} \nonumber \\
&\quad + \begin{pmatrix}
2\alpha k_x^2 & 2\alpha k_xk_y & (\alpha+\beta)k_xk_z \\
2\alpha k_xk_y&2\alpha k_y^2 & (\alpha+\beta)k_yk_z \\
(\alpha+\beta)k_xk_z & (\alpha+\beta)k_yk_z & 2\beta k_z^2
\end{pmatrix},
\end{align}
where $E_{p_x}=E_{p_y}=E_{p_1}$, $E_{p_z}= E_{p_2}$, $\alpha = 2a^2\gamma_{sp}^2/(E_{p_1}-E_s)$ and $\beta = 2a^2\gamma_{sp}^2/(E_{p_2}-E_s)$. The first three terms in Eq.~(\ref{thinfilmHamiltonian}) are the $p$-orbital hopping terms and the last term stems from the $sp$ hybridization. We express the Hamiltonian in terms of the $L_i$ operators to compare with that of the cubic $sp$ case [Eq.~(\ref{cubicsp2})].
\begin{align}\label{thinfilm2}
 \mathcal{H}_p &= E_p-a^2 t_\sigma k^2+a^2(t_\sigma-t_\pi) \sum_ik_i^2L_i^2 \nonumber \\
  &\quad+2\alpha\left(k_x^2+k_y^2 + \frac{\beta}{\alpha} k_z^2\right) 
- 2\alpha k^2L'^2+\frac{(\alpha-\beta)^2k_z^2}{2\alpha}L_z^2,
\end{align}
where $L'=(k_xL_x + k_y L_y + \frac{\alpha+\beta}{2\alpha}k_z L_z)/k$. In the isotropic limit ($\alpha=\beta$), the last two terms becomes $\propto (\vec{L}\cdot \vec{k})^2$ to be identical with Eq.~(\ref{cubicsp2}), while, for $\alpha\ne\beta$, the Hamiltonian becomes anisotropic. 
This Hamiltonian would be useful to analytically model the perpendicular magnetocrystalline anisotropy, which goes far beyond spin-based models like Ref.~\cite{kim2016}.

\subsection{FCC and BCC lattice structure}

As more realistic models, we consider FCC (e.g. Pt) and BCC (e.g. V, Cr, Ta, W) structures. For these structures, we focus on the mechanism (i), which is relevant even without consideration of NNN hoppings. The mechanism (ii) turns out to drive orbital textures in the same form as the cubic lattice [Eqs.~(\ref{cubicsp}), (\ref{pbypd}), (\ref{t2gbypd}), (\ref{egbypd}), and (\ref{peg cubic})] with slightly different coefficients, which are presented in Appendix~\ref{Sec(B):fccbcc2}. In addition, the $t_{2g}e_g$ hybridization, which is forbidden in cubic lattice, is allowed in FCC and BCC lattice structure and drives orbital textures, which appears in higher order in \vec{k} and it is also presented in Appendix~\ref{Sec(B):fccbcc2}.

\subsubsection{FCC structure}
\begin{figure}[]
	\centering
	\includegraphics[width=0.45\textwidth]{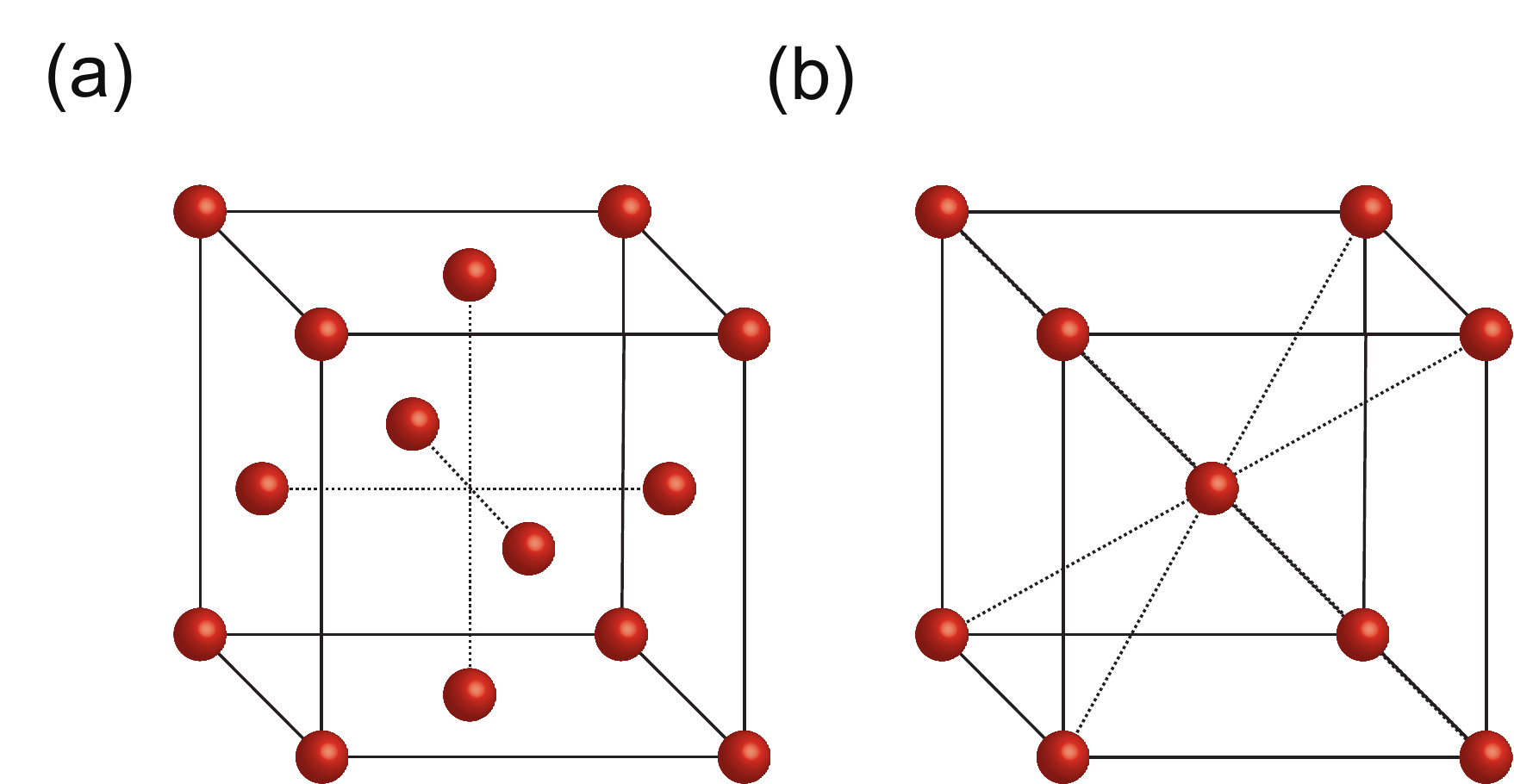}
	\caption{\label{Fig:fccandbcc} (a) Unit cell of FCC and (b) BCC lattice structures. 
	}
\end{figure}

First, we discuss the lattice-driven $p$-orbital and $d$-orbital texture in FCC structure [Fig.~\ref{Fig:fccandbcc}(a)]. The $p$-orbital texture in BCC structure is given in the $(p_x,p_y,p_z)$ basis by
\begin{equation}
\mathcal{H}_p  = E_p - a^2(3t_\pi + t_\sigma)k^2 -2a^2(t_\sigma-t_\pi)\begin{pmatrix}
\frac{k_x^2}{2} & k_xk_y & k_xk_z\\
k_xk_y & \frac{k_y^2}{2} & k_yk_z\\
k_xk_z&k_yk_z& \frac{k_z^2}{2}
\end{pmatrix},
\end{equation}
which can be expressed in terms of $L_i$ operators as
\begin{equation}
\mathcal{H}_p = E_p -2a^2(t_\sigma+t_\pi)k^2 
+a^2(t_\sigma-t_\pi)\sum_{i,j}k_ik_j \{L_i,L_j\}.
\end{equation}

Next, we show the Hamiltonian for the $t_{2g}$ orbitals in the ($d_{yz}$, $d_{zx},$ $d_{xy}$) basis:
\begin{align}
\mathcal{H}_{t_{2g}} &= E_{t_{2g}}-\frac{a^2}{2}(2V_{dd\pi}+3V_{dd\delta}+3V_{dd\sigma})k^2 \nonumber \\
 &\quad+\frac{a^2(V_{dd\delta}-V_{dd\pi})}{2} \begin{pmatrix}
\xi k_x^2 & 4k_xk_y & 4k_xk_z \\
4k_xk_y & \xi k_y^2 & 4 k_yk_z \\
4 k_xk_z& 4k_yk_z & \xi k_z^2
\end{pmatrix}, 
\end{align}
where $\xi = (2V_{dd\pi}+V_{dd\delta}-3V_{dd\sigma})/(V_{dd\pi}-V_{dd\delta})$. This Hamiltonian is expressed in terms of $L_i$ operators as
\begin{align}
\mathcal{H}_{t_{2g}} &= E_{t_{2g}}-\frac{a^2}{2}(3V_{dd\pi}+3V_{dd\sigma}+2V_{dd\delta})k^2  \nonumber \\
&\quad+\frac{a^2\xi(V_{dd\pi}-V_{dd\delta})}{2}\sum_ik_i^2L_i^2  
\nonumber\\
&\quad+a^2(V_{dd\pi}-V_{dd\delta})\sum_{i,j(\neq i)}k_ik_j \{L_i,L_j\}.
\end{align}

Finally, the $e_g$-orbital texture in the $d_{x^2-y^2},d_{z^2}$ basis is given by
\begin{align}
\mathcal{H}_{e_g} &= E_{e_g}-\frac{a^2}{2}(4V_{dd\pi}+3V_{dd\delta}+V_{dd\sigma})k^2 \nonumber \\
 &\quad +\mu \begin{pmatrix}
k^2-3k_z^2 & -\sqrt{3}(k_x^2-k_y^2)\\
-\sqrt{3}(k_x^2-k_y^2) & -k^2+3k_z^2
\end{pmatrix},
\end{align}
where $\mu =a^2(3V_{dd\delta}+V_{dd\sigma}-4V_{dd\pi})/8$. We can see that it has the same $e_g$ orbital texture structure as that in cubic lattice structure [Eq.~(\ref{3degtexture})].

\subsubsection{BCC structure}

Next, we demonstrate $p$- and $d$-orbital textures in the BCC structure [Fig.~\ref{Fig:fccandbcc}(b)]. First, for a system with $p$ orbitals only, the Hamiltonian is given by
\begin{equation}
\mathcal{H}_p =E_{p}-\frac{4a^2(2t_\pi+t_\sigma)k^2}{3}
 + \frac{2a^2(t_\pi-t_\sigma)}{3}\begin{pmatrix}
0&k_xk_y & k_xk_z\\
k_xk_y&0&k_yk_z\\
k_xk_z&k_yk_z & 0
\end{pmatrix},
\end{equation}
which can be expressed in terms of $L_i$ operators as
\begin{equation}
\mathcal{H}_p = E_p - \frac{4a^2(2t_\pi+t_\sigma)}{3}k^2  +
\frac{a^2(t_\sigma-t_\pi)}{3}\sum_{i,j(=\neq i)}k_ik_j \{L_i,L_j\}.
\end{equation}

The Hamiltonian of a system with $t_{2g}$ orbitals only is given by
\begin{align}
\mathcal{H}_{t_{2g}}&= E_{t_{2g}}-\frac{4a^2}{9}(2V_{dd\pi}+4V_{dd\delta}+3V_{dd\sigma})k^2 \nonumber \\ 
&\quad+\frac{8a^2}{9}(V_{dd\pi}+2V_{dd\delta}-3V_{dd\sigma}) \begin{pmatrix}
0& k_xk_y & k_xk_z \\
k_xk_y & 0& k_yk_z \\
k_xk_z&k_yk_z&0
\end{pmatrix},
\end{align}
which can be expressed as
\begin{align}
\mathcal{H}_{t_{2g}} &= E_{t_{2g}}-\frac{4a^2}{9}(2V_{dd\pi}+4V_{dd\delta}+3V_{dd\sigma})k^2 +\nonumber \\ 
&\quad-\frac{4a^2}{9}(V_{dd\pi}+2V_{dd\delta}-3V_{dd\sigma})\sum_{i,j=1,i\neq j}^3k_ik_j \{L_i,L_j\}.
\end{align}
For the BCC case, $p$ and $t_{2g}$ orbitals already have the orbital textures, but there exists some planes where there are no orbital texture. For example, at $k_z = 0$ plane, the Fermi surface is of the form of Fig.~\ref{Fig:otsquare}(a) with $\pi/4$ rotation (degeneracy exists in $k_x=0$ and $k_y=0$ lines). Same as in the square lattice model, hybridization between orbitals with different $l$'s ($sp$, $pd$ hybridizations) opens gaps and play important roles around this points. As mentioned above, the orbital textures formed by the mechanism (ii) has same form as that of the cubic lattice which is in Appendix~\ref{Sec(B):fccbcc2}. Lastly, $e_g$ orbital has no lattice driven orbital texture in BCC structure.

In summary, for the BCC and FCC structures, the mechanism (i) and (ii) work together, forming orbital textures beyond the simple $r$-$t$ model.


\section{Application: Orbital hall conductivity \label{Sec:discussion}}


The OHE is arguably the most representative phenomenon originating from orbital textures. The strength of the OHE may be quantified by the orbital Hall conductivity (OHC), which measures the amount of the orbital Hall current generated by an applied electric field. In this section, we use some models derived above to demonstrate that the OHC may vary significantly, depending on the type of the orbital texture. As illustrative examples, we investigate the OHC for square lattices with $sp$ or $pt_{2g}$ orbitals where the $p$-orbital texture changes between the $r$-$t$ and Dressehalus type [Eqs.~(\ref{squareporbital}) and (\ref{pdHamiltonian})] depending on the difference of the on-site energies. The OHC is obtained by the Kubo formula.
\begin{align}
\sigma^z_{yx}&= \sum_{n,\vec{k}}f_{n,\vec{k}}\sigma_{yx,n\vec{k}}^z,\\
\sigma_{yx,\vec{k},n}^z&= \frac{\hbar e^2}{V}\Im \sum_{n'(\ne n)}\frac{\bra{n,\vec{k}}\{L_z,v_y\}\ket{n',\vec{k}}\bra{n',\vec{k}}v_x\ket{n,\vec{k}}}{(E_{n,\vec{k}}-E_{n',\vec{k}})^2},
\end{align}
where $n$ is the band index, $V$ is the volume of the system, $f_{n,\vec{k}}$ is the Fermi-Dirac distribution, $\sigma_{yx,n\vec{k}}^z$ is the contribution to the OHC from the the $(n,\vec{k})$ state, $\vec{v}=(1/\hbar)\partial_\vec{k}\mathcal{H}$ is the velocity operator, and $E_{n,\vec{k}}$ is the energy eigenvalue of $\mathcal{H}$ with respect to the $(n,\vec{k})$ state.

After some algebra, the OHE for the outer band is
\begin{eqnarray}
\sigma^z_{yx,\rm outer,\vec{k}} =\frac{2e^2a^4 k^2 k_y^2 A(t_\pi^2-t_\sigma^2)}{\hbar V(E_1-E_2)^3},
\end{eqnarray}
where $A =4\gamma_{sp}^2a^2/(E_p-E_s)$ for the $sp$ system and $A=4V_{pd\pi}^2a^2/(E_p-E_{t_{2g}})$ for the $pt_{2g}$ system and $E_1$ and $E_2$ refer to the energy of the inner and the outer $p$ bands, respectively. Here OHC is calculated up to second order in hybridization energies ($\gamma_{sp}$ or $V_{pd\pi}$). Note that the sign of OHC depends on the difference of $E_p$ and $E_s$ (or $E_{t_{2g}}$). While the previous models with phenomenological orbital texture parameter~\cite{park2022} cannot give an insight on the sign of OHC, our formalism clearly shows its direct connection to microscopic parameters. In addition, we remark that the sign of $A$ determines whether the orbital texture is in the $r$-$t$ type or the orbital Dresselhaus type and thus the sign of OHC depends on the geometrical type of the orbital texture. We believe that our formalism would shed light on the negative OHC reported in a previous work~\cite{baek2021}.

\section{Summary\label{Sec:summary}}

In this paper, we microscopically derive the orbital texture Hamiltonian for various cases by considering two mechanisms: the lattice structure and the orbital hybridization with other orbitals with $l$. 
In many realistic materials, the lattice structure already drives system to have an orbital texture even without hybridization between orbitals with different $l$'s. The orbital hybridization by the two mechanisms plays important role near degeneracy points and form an orbital texture. Our calculations show that a bilayer structure exhibits hidden orbital Rashba states which may explain previous experimental observations~\cite{tsai2018,park2018}. Our formalism also sheds light on the microscopic origin of the qualitatively different behaviors of the OHC (such as its sign) depending on systems. Our formalism will be useful for constructing orbital texture models for many situations to describe diverse orbital physics, including the magnetocrystalline anisotropy and the orbital transport phenomena.

\begin{acknowledgments}
We thanks J. Sohn and D. Jo for fruitful discussions. S. H. and H.-W. L. were supported by the Samsung Science and Technology Foundation (BA-1501-51). K.-W. K. was supported by the KIST Institutional Programs (2E31541, 2E31543) and the National Research Foundation (NRF) of Korea (2020R1C1C1012664).
\end{acknowledgments}

\begin{appendix}
\section{Review of L\"owdin downfolding\label{Sec(A):Lowdin}}

Here, we briefly review the L\"owdin downfolding technique based on Refs.~\cite{lowdin1951,winkler2003}. We also present a simple example, a square lattice with $sp$ system, and explicitly show how to perform the downfolding for this case. The L\"owdin downfolding is a technique which block-diagonalizes the total Hamiltonian up to a desired order. When the interaction between a subspace that we are interested in and the others is weak, one can perturbatively block-diagonalizes the Hamiltonian. Then one can obtain the effective subspace Hamiltonian by taking the corresponding block. For example, in main text we apply L\"owdin downfolding to the $sp$ Hamiltonian, making effective Hamiltonian of $p$-character bands. 

We divide the total Hamiltonian as two parts , $H = H^0 + \lambda H'$. For simplicity, we assume $H^0$ is diagonal Hamiltonian and we know its eigenstates and eigenvalues. $\lambda H'$ describes interaction term between the block that we are interested in and the other blocks. We assume $\lambda H'$ is weak and we consider this term as perturbation. 
\begin{equation}
	H = H^0 + \lambda H'.
\end{equation}
Next, we perform a unitary transformation to make the total Hamiltonian block diagonal up to desired order.
\begin{equation}
	\tilde{H} = e^{-S}He^{S}  
	= H + [H,S] + \frac{1}{2!}[[H,S],S]+\cdots ,
\end{equation}
where $S$ is anti-hermitian, $S^\dagger = -S$, so that $e^S$ is unitary. For $\lambda=0$, $S=0$ diagonalizes $H$ so that $S$ is at least first order in $\lambda$. We expand the $S$ as $S = \lambda S^{(1)}+ \lambda^2 S^{(2)}+\cdots$ and determine the $S^{(n)}$ by making non-diagonal terms in $\tilde{H}$ become zero. Then up to $\lambda^2$ order,
\begin{align}
	\tilde{H} &= H^0 +\lambda H' + [H^0 + \lambda H' ,\lambda S^{(1)} ] \\
	&\quad + \frac{1}{2}[[H^0,\lambda S^{(1)}],\lambda S^{(1)}] + [H^0,\lambda^2 S^{(2)}]+\mathcal{O}(\lambda^3).
\end{align}
We impose conditions for $S^{(1)}$, $S^{(2)}$ as,
\begin{align}
	\lambda H' + [H^0,\lambda S^{(1)}]&= 0 \label{s1condition} \\
	S^{(2)}&=0.
\end{align}
which are satisfied by
\begin{equation}
	S^{(1)}_{mn} = -\frac{H'_{mn}}{E_m-E_n},~S^{(2)}_{mn}=0,
\end{equation}
where $E_m$ and $E_n$ are eigenvalues of $H_0$ with respect to $m$ and $n$ states, respectively. Then, the transformed Hamiltonian is given by
\begin{equation}
	\tilde{H} 
	=  H^0  +  \frac{\lambda^2}{2}[H',S^{(1)}]+ \mathcal{O}(\lambda^3),
\end{equation}
The component-wise expression of $\tilde{H}_{mn}$ is given by
\begin{equation}
	\tilde{H}_{mn} = H^0_{mn} -\frac{\lambda^2}{2}\sum_l H'_{ml}H'_{ln}\left[ \frac{1}{E_l-E_n}+\frac{1}{E_l-E_m}\right].\label{downfolding}
\end{equation}
Note that the block-off-diagonal components $H'$ are eliminated. The cost of the cancellation is the appearence of the second order block-diagonal corrections, which give the effective Hamiltonian for the desired block.

As an illustrative example, we apply the L\"owdin downfolding technique to the square lattice $sp$ model. We start from the Hamiltonian [Eq.~(\ref{sptightbinding})] expanded up to $k^2$ order
\begin{widetext}
\begin{equation}
	\mathcal{H} = \begin{pmatrix}
		E_s-a^2t_sk^2& 0 & 0 \\
		0 & E_p-a^2(t_\sigma k_x^2+t_\pi k_y^2) & 0\\
		0 & 0& E_p-a^2(t_\pi k_x^2+t_\sigma k_y^2) 
	\end{pmatrix}+\begin{pmatrix}
	0& 2ia\gamma_{sp}k_x & 2ia\gamma_{sp}k_y \\
	-2ia\gamma_{sp}k_x &0 & 0\\
	-2ia\gamma_{sp}k_y & 0& 0
\end{pmatrix},
\end{equation}
\end{widetext}
where the first term corresponds to $H_0$ and the second term corresponds to $\lambda H'$. Therefore, we use Eq.~(\ref{downfolding}) to immediately obtain the effective Hamiltonian for the $p$ block as
\begin{align}
	\tilde{\mathcal{H}}_p &= \mathcal{H}^0_p + \frac{1}{E_p-E_s}\begin{pmatrix}
		-2ia\gamma_{sp}k_x\\
		-2ia\gamma_{sp}k_y
	\end{pmatrix} \begin{pmatrix}
		2ia\gamma_{sp}k_x & 2ia\gamma_{sp}k_y
	\end{pmatrix} \nonumber \\
	&= \mathcal{H}_p^0 + \frac{4a^2\gamma_{sp}^2}{E_p-E_s} \begin{pmatrix}
		k_x^2 & k_xk_y \\
		k_xk_y & k_y^2
	\end{pmatrix}.
\end{align}
Here, we have used $\mathcal{H}^0_{p_ip_i}-\mathcal{H}^0_{ss} \approx E_p-E_s$ where $i=x,y $ approximation assuming $E_p-E_s$ is large. This gives Eq.~(\ref{squareporbital}) in the main text.

\section{Hybridized states effect on the OAM operators\label{Sec(A):pr+is}}

In Appendix~\ref{Sec(A):Lowdin}, we have focused on the effective Hamiltonian $\tilde{H}$ after a unitary transformation. It is notable that the unitary transform may alter the eigenstates. For instance, in the 
triangular $sp$ model, the eigenstates of the Eq.~(\ref{triangularsp}) are given by $\tilde{p_r}=\cos(\theta/2)\ket{p_r} + i \sin(\theta/2)\ket{s} , \tilde{s}=\cos(\theta/2)\ket{s} + i \sin(\theta/2)\ket{p_r}, p_{t1},p_{t2}$. That is, the $p_r$ orbital carries imaginary $s$-orbital character and vice-versa while tangential orbitals remain same. Accordingly, 
the physical operators written in this basis can be different from that written in the pristine $s$ and $p$ states. 
Here, we investigate how the OAM operators change under the hybridized states. Under the unitary transformation operator $U=\ket{\tilde{s}}\bra{s}+ \ket{\tilde{p_r}}\bra{p_r}+ \ket{t_1}\bra{t_1}+\ket{t_2}\bra{t_2}$, the OAM operators are transformed as follows.
\begin{align}
	U L_r U^\dagger &= L_r \\
	U L_{t_2}U^\dagger &= \begin{pmatrix}
		0&0&\sin(\frac{\theta}{2})& 0 \\
		0&0&i\cos(\frac{\theta}{2}) &0 \\
		\sin(\frac{\theta}{2})&-i\cos(\frac{\theta}{2})& 0 &0 \\
		0&0&0&0
	\end{pmatrix}.
\end{align}
$L_{t1}$ transforms similar to the $L_{t2}$ case only interchanging component of the $s,p_r \leftrightarrow p_{t1}$ to the $s,p_r \leftrightarrow p_{t2}$. There are few remarks. First, $L_r$ does not change since $sp$ hybridization does not affect tangential orbitals.\footnote{Note that nonzero $L_r$ is generated by an imaginary mixture of the tangential $p$ orbitals.} Second, if we project $L_{t2}$ operator onto $p$-orbital subspace like in Ref.~\cite{han2022}, then it becomes usual OAM operator in $p$-orbital space since $\cos(\theta/2) \approx 1 $ up to first order in $1/(E_p-E_s)$. Finally, there exists nontrivial off-diagonal term between $\tilde{s}-p_t$ orbital since $\tilde{s}$ orbital carries imaginary $p_r$ orbital character. This term is proportional to $\sin(\theta/2)\sim 1/(E_p-E_s)$ which may give a nonnegligible contribution. Therefore, to fully describe the $p$-orbital dynamics using the effective downfolding Hamiltonian, the $s$-orbital degree of freedom should be considered. Based on the $L$ operators constructed above, $\{L_i,L_j\}$ can be calculated and result in similar conclusions with the $L$ operators. 

\section{Effects of the mechanism (ii) for FCC and BCC structures\label{Sec(B):fccbcc2}}


First, we show corrections to the effective Hamiltonian of $p$-character bands by hybridization with other orbitals. The correction term by the $sp$ hybridization is given by
\begin{align}
	\Delta \mathcal{H}_{p}^{sp} &= \frac{\alpha}{E_p-E_s}\begin{pmatrix}
		k_x^2 & k_xk_y & k_xk_z \\
		k_xk_y & k_y^2 & k_yk_z \\
		k_xk_z & k_yk_z & k_z^2
	\end{pmatrix},  \\
	\alpha&=\begin{cases}
		32a^2\gamma_{sp}^2&\text{ for FCC},\\
		\frac{64a^2}{3}\gamma_{sp}^2&\text{ for BCC}.
	\end{cases}
\end{align}
Next, the correction term by the $pt_{2g}$ hybridization is given by
\begin{align}
	\Delta \mathcal{H}_{p}^{pt_{2g}} &= \frac{\beta}{E_p-E_{t_{2g}}} \begin{pmatrix}
		k_y^2+k_z^2 & k_xk_y & k_xk_z \\
		k_xk_y & k_x^2+k_z^2 & k_yk_z \\
		k_xk_z & k_yk_z & k_x^2+k_y^2
	\end{pmatrix},\\
	\beta&=\begin{cases}
	2a^2(2V_{pd\pi}+\sqrt{3}V_{pd\sigma})^2&\text{ for FCC},\\
	\frac{64a^2}{81}(\sqrt{3}V_{pd\pi}+3V_{pd\sigma})^2&\text{ for BCC}.
\end{cases}
\end{align}
The correction term by the $pe_g$ hybridization is given by
\begin{align}
	\Delta \mathcal{H}_{p}^{pe_g} &= \frac{\delta}{E_p-E_{e_g}} \begin{pmatrix}
		2k_x^2 & -k_xk_y & -k_xk_z \\
		-k_xk_y & 2k_y^2 & -k_yk_z \\
		-k_xk_z & -k_yk_z & 2k_z^2
	\end{pmatrix}, \\
	\delta&=
	\begin{cases}
	a^2(2\sqrt{3}V_{pd\pi}+V_{pd\sigma})^2&\text{ for FCC},\\
	\frac{128a^2}{9}V_{pd\pi}^2&\text{ for BCC}.
	\end{cases}
\end{align}

Next, we show corrections to the effective Hamiltonian of $d$-character bands by hybridization with $p$ orbitals. The correction term tp the effective Hamiltonian of $t_{2g}$ orbital character band is given by
\begin{align}
	\Delta \mathcal{H}_{t_{2g}}^{pt_{2g}} &= \frac{\beta}{E_{t_{2g}}-E_p} \begin{pmatrix}
		k_y^2+k_z^2 & k_xk_y & k_xk_z \\
		k_xk_y & k_x^2+k_z^2 & k_yk_z \\
		k_xk_z & k_yk_z & k_x^2+k_y^2
	\end{pmatrix},
\end{align}
where $\beta$ is the same as above. Lastly, the correction term to the $e_g$ orbital Hamiltonian is
\begin{align}
	\Delta \mathcal{H}_{e_g}^{pe_g}&=\frac{\xi}{E_{e_g}-E_p}\left[2k^2 - \begin{pmatrix}
		-k^2+3k_z^2& \sqrt{3}(k_x^2-k_y^2)\\
		\sqrt{3}(k_x^2-k_y^2) &k^2-3k_z^2 
	\end{pmatrix}\right],\\
	\xi&=\begin{cases}
		\frac{a^2}{2}(2\sqrt{3}V_{pd\pi}+V_{pd\sigma})^2&\text{ for FCC},\\
		\frac{64a^2}{9}V_{pd\pi}^2&\text{ for BCC}.
	\end{cases}
\end{align}
For $t_{2g}e_g$ hybridization term in $(d_{yz}$, $d_{zx}$, $d_{xy}$, $d_{x^2-y^2}$, $d_{z^2}$) basis is given,
\begin{align}
	\Delta \mathcal{H}^{t_{2g}e_g} &= \rho\begin{pmatrix}
	0&0&0&\sqrt{3}k_yk_z&-k_yk_z \\
	0&0&0&-\sqrt{3}k_xk_z & -k_xk_z \\
	0&0&0&0&2k_xk_y \\
	\sqrt{3}k_yk_z&-\sqrt{3}k_xk_z&0&0&0 \\
	-k_yk_z&-k_xk_z&2k_xk_y &0&0
\end{pmatrix}, \\
 \rho&= \begin{cases}
        \frac{\sqrt{3}a^2}{2}(V_{dd\sigma}-V_{dd\delta})&\text{ for FCC},\\
		\frac{8a^2}{3\sqrt{3}}(V_{dd\pi}-V_{dd\delta})&\text{ for BCC}.     
 \end{cases}
\end{align}

\end{appendix}

\end{document}